\documentclass[11pt,a4paper,onecolumn,notitlepage]{article}
\let\Horig\H

\let\Lorig\L
\usepackage{CJKutf8}
\usepackage{cmap} 
\usepackage[pdfpagemode=None,
pagebackref=true,
bookmarksopen=false,
hyperindex=true,
colorlinks=true,
linkcolor=blue,
citecolor=blue,
pdftitle={Towards postquantum Va\u{\i}nberg--Br\`{e}gman relative entropies},
pdfauthor={Ryshard-Pavel Kostecki}]{hyperref}

\usepackage[T2A,T1]{fontenc} 
\usepackage{amsfonts} 
\usepackage{amsmath} 
\usepackage{amssymb} 
\usepackage{amsthm} 
\usepackage{upgreek} 
\usepackage{mathrsfs} 
\usepackage{combelow}
\usepackage{empheq}
\usepackage{enumitem}
\usepackage{scalerel}
\usepackage{roundrulemod}
\newcommand{\sgn}{\mathrm{sgn}}   
\newcommand{\rpkmark}[1]{#1}
\newcommand{\jordan}{{\,\scalebox{0.8}{$\odot$}\,}}
\newcommand{\re}{\mathrm{re}\,}   
\newcommand{\X}{\mathcal{X}}   
\newcommand{\DDD}{\mathfrak{D}}     
\newcommand{\inver}{{\horizrule\,\vertrule{\mkern 1.6mu}}}
\renewcommand{\mid}{\,\st}
\newcommand{\st}{\boldsymbol{:}\,}
\newcommand{\pclg}{\Gamma^\mathrm{G}}
\newcommand{\pcl}{\Gamma}
\newcommand{\orlicz}{\Upsilon}%
\newcommand{\LPPP}{\overleftarrow{\mathfrak{P}}}     \newcommand{\RPPP}{\overrightarrow{\mathfrak{P}}}     
\newcommand{\BBB}{\mathfrak{B}}   
\newcommand{\DF}{\DDD^{\mathrm{F}}}
\newcommand{\DG}{\DDD^{\mathrm{G}}}
\renewcommand{\H}{\mathcal{H}} 
\newcommand{\N}{\mathcal{N}}   
\newcommand{\NN}{\mathbb{N}} 
\newcommand{\RR}{\mathbb{R}} 
\newcommand{\CC}{\mathbb{C}} 
\newcommand{\KK}{\mathbb{K}} 
\newcommand{\II}{\mathbb{I}} 
\newcommand{\dd}{\mathrm{d}}  
\newcommand{\gbold}{\mathbf{g}}
\newcommand{\T}{\mathbf{T}}          
\newcommand{\transport}{\mathbf{t}}   
\newcommand{\lfdual}{\mathbf{F}}   
\newcommand{\efd}{\mathrm{efd}}   
\newcommand{\ra}{\rightarrow}      
\newcommand{\limp}{\Rightarrow}    
\newcommand{\BH}{{\BBB(\H)}}           
\newcommand{\tr}{\mathrm{tr}}     
\newcommand{\grad}{\mathrm{grad}} 
\newcommand{\INT}{\mathrm{int}} 
\newcommand{\iso}{\cong}           

\newcommand{\duality}[1]{\left[\!\left[#1\right]\!\right]}
\newcommand*{\n}[1]{{\left|\!\left|{#1}\right|\!\right|}} 
\newcommand*{\arginff}[2]{\arginf_{#1}\left\{#2\right\}}

\newcommand*{\intefd}[1]{\INT(\efd(#1))}
\newcommand*{\df}[1]{\textit{\textbf{#1}}}
\newcommand*{\cytat}[1]{\guillemotleft{#1}\guillemotright}
\newcommand*{\ab}[1]{{\left|{#1}\right|}} 

\DeclareMathOperator*{\arginf}{\arg\inf}

\DeclareMathAlphabet{\mathpzc}{OT1}{pzc}{m}{it}
\newcommand{\xx}{\mathpzc{x}}      

\newtheorem{lemma}{Lemma}[section]

\newtheorem{proposition}[lemma]{Proposition}
\newtheorem{definition}[lemma]{Definition}
\newtheorem{corollary}[lemma]{Corollary}

	\theoremstyle{definition}%
	\newtheorem{remark}[lemma]{Remark}%

\newcommand{\MMM}{\mathscr{M}}     
\newcommand{\Orlicz}{\Upsilon}
\newcommand{\young}{\mathbf{Y}}
\newcommand{\DIFF}{{\mathrm{C}^1}} 
\newcommand{\NFUN}{\mathrm{N}} 
\newcommand{\SC}{\mathrm{SC}} 
\newcommand{\type}{\widetilde{\mathrm{type}}}
\newcommand{\Isfi}{I$^{\,\text{s.f.}}_\infty$} 
\newcommand{\koethe}{\times}
\newcommand*{\rearr}[2]{{{#1}^{#2}}} 

\usepackage{geometry}
\usepackage{bibspacing}

\renewcommand*{\backref}[1]{}
\renewcommand*{\backrefalt}[4]{%
  \ifcase #1 %
    \relax
  \or
    $\uparrow$~#2.
  \else
    $\uparrow$~#2.
  \fi%
}

\begin{document}

\title{\begin{changemargin}{-1cm}{-1cm}\begin{center}{\vskip -1.5cm}{\huge Towards postquantum Va\u{\i}nberg--Br\`{e}gman relative entropies}\end{center}\end{changemargin}}

\author{{\Large Ryshard-Pavel Kostecki}\\
{\small }\\
{\small \textit{Research Center for Quantum Information, Slovak Academy of Sciences}}\\
{\small\textit{D\'{u}bravsk\'{a} cesta 9, 845 11 Bratislava, Slovakia}}\\
{\small\texttt{kostecki@fuw.edu.pl}}\\
{\small\texttt{\href{https://www.fuw.edu.pl/~kostecki}{www.fuw.edu.pl/$\sim$kostecki}}}
}
\date{27 June 2026}
\maketitle      
\thispagestyle{empty}       
\begin{abstract}
\noindent 
We develop a new approach to construction of the Va\u{\i}nberg--Br\`{e}gman relative entropies over nonreflexive Banach spaces, based on nonlinear embeddings into reflexive Banach spaces. We apply it to derive some new families of Va\u{\i}nberg--Br\`{e}gman relative entropies over some radially compact base normed spaces in spectral duality, and to establish their basic properties. In particular, we prove (left and right) generalised pythagorean theorem and norm-to-norm continuity of the left entropic projections for a family of Va\u{\i}nberg--Br\`{e}gman relative entropies induced on preduals of any W$^*$-algebras (resp., semifinite JBW-algebras) using Mazur maps into noncommutative (resp., nonassociative) $L_p$ spaces, on preduals of semifinite W$^*$-algebras using Kaczmarz maps into noncommutative Orlicz spaces, and on (resp., positive parts of) unit spheres of commutative $L_1$ spaces (resp., trace class operators) using Lozanovski\u{\i} factorisation maps. We also prove left generalised pythagorean theorem for a family of Va\u{\i}nberg--Br\`{e}gman relative entropies over preduals of generalised spin factors. Additionally, we characterise strict convexity, Gateaux differentiability, Radon--Riesz--Shmul'yan property, and reflexivity of the Morse--Transue--Nakano and Orlicz norms on noncommutative Orlicz spaces, establish Lipschitz--H\"{o}lder continuity of the nonassociative Mazur map on positive parts of unit balls, and introduce a new class of $L_p$ spaces over spectrally dual order unit spaces.
\end{abstract}

\section{Introduction}

For any set $W$, $D:W\times W\ra[0,\infty]$ will be called an \df{information} on $W$ (and $-D$ will be called a \df{relative entropy}\footnote{Cf. \cytat{information is the negative of the quantity (...) defined as entropy} \cite[p. 76]{Wiener:1948}. This convention agrees with: R\'{e}nyi's \cytat{measure of the amount of information} \cite[p. 554]{Renyi:1961}, Umegaki's definition of \cytat{information} on state spaces of type I W$^*$-algebras \cite[Def. 1]{Umegaki:1961}, Csisz\'{a}r's definition of \cytat{relative information} \cite[p. 86]{Csiszar:1963}, as well as with the sign, ordering, and naming conventions used in \cite{Bratteli:Robinson:1979:1981}.} on $W$) if{}f $D(x,y)=0$ $\iff$ $x=y$ $\forall x,y\in W$ (cf. \textup{\cite[p. 1019]{Bregman:1966}} \textup{\cite[p. 794]{Eguchi:1983}} \textup{\cite[p. 161]{Csiszar:1995}}). If $\varnothing\neq K\subseteq W\ni x$, and $\arginff{y\in K}{D(y,x)}$ (resp., $\arginff{y\in K}{D(x,y)}$) is a singleton set, then we will denote the element of this set by $\LPPP^D_K(x)$ (resp., $\RPPP^D_K(x)$), the map $x\mapsto\LPPP^D_K(x)$ \cite[p. 32]{Sanov:1957} \cite[Ch. 3.2]{Kullback:1959} (resp., $x\mapsto\RPPP^D_K(x)$ \cite[Eqn. (16)]{Chencov:1968}) will be called a \df{left} (resp., \df{right}) \df{$D$-projection} of $x$ onto $K$, while $K$ will be called a \df{left} (resp., \df{right}) \df{$D$-Chebysh\"{e}v} set. 

In this paper we are concerned with a construction of the family of the extended Va\u{\i}nberg--Br\`{e}gman relative entropies $-D_{\ell,\Psi}$ with $W$ given by subsets of nonreflexive Banach spaces (in general) and subsets of radially compact base normed spaces in spectral duality (in particular).

If $B$ is a base of a radially compact base normed space $(V,\n{\cdot}_V)$, then it admits a convex theoretic generalisation of a probability theory\footnote{Called also a ``generalised probability theory'' or a ``postquantum theory'' (while the adjective `postquantum' has a somewhat questionable flavour, its virtue is in concision; we use it exclusively to refer to the setting of radially compact base normed spaces).} \cite{Gunson:1967,Mielnik:1969,Davies:Lewis:1970,Ludwig:1970,Gudder:1973}, allowing to interpret any $W\subseteq B$ as a set of `information states', and to interpret the pair $(W,D)$ as an `information geometry'. Furthermore, if $(V,\n{\cdot}_V)$ is in spectral duality, then it admits a spectral theory allowing to develop absolute integration theory such that $(V,\n{\cdot}_V)$ can be identified with a suitably defined $L_1$ space, so the `information states' can be identified with (the normalised densities of) absolute integrals (see Remark \ref{remark.postquantum.Lp} for more details). In such case the information geometry $(W,D)$ can be considered in terms of Chencov's programme of geometrostatistics \cite{Chencov:1964,Chencov:1965,Chencov:1968,Chencov:1972,Morozova:Chencov:1989,Morozova:Chencov:1991}, aiming at characterisation of the properties of various types of statistical inferences by means of the geometric structures on the sets of (densities of) absolute integrals.\footnote{Relative entropies on radially compact base normed spaces in spectral duality were first considered in \cite[\S3]{Tikhonov:1990} and \cite[\S4]{Zanzinger:1995} (cf. also \cite[p. 360]{Zanzinger:1998}).}

Given a strictly convex and differentiable function $\Psi:\RR^n\ra\RR$, the Va\u{\i}nberg--Br\`{e}gman information $D_\Psi$, defined by \eqref{eqn.Vainberg.Bregman.D.Psi}, quantifies the difference between $\Psi$ and the first order Taylor expansion of $\Psi$. The key feature of $D_\Psi$ is the (right and left) generalisation of a pythagorean theorem\footnote{The latter is recovered for $\Psi(x)=\frac{1}{2}\sum_{i=1}^nx_i^2$.}, with the orthogonal projection on the convex and closed subset $K$ of $\RR^n$ replaced by the (right and left) $D_\Psi$-projection. For a nonempty convex closed $C\subseteq M\subseteq\RR^n$ and $\varnothing\neq\widetilde{C}\subseteq M\subseteq\RR^n$ with convex closed $(\grad\Psi)(\widetilde{C})$, $D_\Psi$ exhibits \cite[Lem. 1]{Bregman:1966} 
\begin{equation}
D_\Psi(x,\LPPP^{D_\Psi}_C(y))+D_\Psi(\LPPP^{D_\Psi}_C(y),y)\leq D_\Psi(x,y)\;\forall(x,y)\in C\times M,
\label{eqn.gen.pyth.left}
\end{equation}
with $\leq$ replaced by $=$ if $C$ is affine, and \cite[Prop. 4.11]{MartinMarquez:Reich:Sabach:2012}\footnote{First instance of this inequality, together with its interpretation as a \cytat{nonsymmetrical analogue of the theorem of Pythagoras}, was established in \cite[Thm. 1]{Chencov:1968} for the Kullback--Leibler information, i.e. for $D_\Psi$ with $\Psi(x_i)=\sum_{i=1}^n(x_i\log(x_i)-x_i)$. The special case of \eqref{eqn.gen.pyth.left} for the Kullback--Leibler information was first considered implicitly in \cite[p. 1021]{Bregman:1966} and explicitly in \cite[Thm. 2.2]{Csiszar:1975}.}
\begin{equation}
D_\Psi(x,\RPPP^{D_\Psi}_{\widetilde{C}}(x))+D_\Psi(\RPPP^{D_\Psi}_{\widetilde{C}}(x),y)\leq D_\Psi(x,y)\;\forall(x,y)\in M\times \widetilde{C},
\label{eqn.gen.pyth.right}
\end{equation}
with $\leq$ replaced by $=$ if $(\grad\Psi)(\widetilde{C})$ is affine. This property can be interpreted as an additive decomposition of information about ``data'' into information about  ``signal'' and information about ``noise''. It is a fundamental feature of $D_\Psi$, characterising $\LPPP^{D_\Psi}_C$ and $\RPPP^{D_\Psi}_{\widetilde{C}}$ (see Section \ref{section.bregman.reflexive}).

So far there have been three different approaches generalising the setting of $D_\Psi$ on $\RR^n$ to more general scenarios: 1) measure/trace theoretic approach; 2) reflexive Banach space approach; 3) finite dimensional $\mathrm{C}^\infty$-differentiable information geometric approach. None of them is able to address the case when $W$ is (a subset of) a predual of an arbitrary W$^*$-algebra or of an arbitrary semifinite JBW-algebra. We develop here a new approach that solves this problem in a quite general way, combining the reflexive Banach space approach with the suitable Banach space generalisation of the nonlinear embeddings that are featured in the information geometric approach. In what follows we will first explain the reasons for the preference of the reflexive approach over the measure/trace theoretic approach, then we will present the idea of the nonlinear embeddings in the original information geometric context (also in order to show the backwards compatibility with the finite dimensional case), and then we will present the main aspects and features of our construction, providing the short guide to the rest of this paper.

\subsection{Va\u{\i}nberg--Br\`{e}gman vs Brunk--Ewing--Utz}

Let $d,m,n\in\NN$. Given a strictly convex and differentiable function $\Psi:\RR^n\ra\RR$ (or $\Psi:M\ra\RR$ with convex $M\subseteq\RR^n$), there are two approaches to construction of an information functional on $\RR^d$ encoding the first order Taylor expansion of $\Psi$ (together with its further use in convex optimisation problems): one going back to Va\u{\i}nberg \cite[Eqn. (8.5)]{Vainberg:1956} and Br\`{e}gman \cite[p. 1021]{Bregman:1966} (=\cite[Eqn. (2.1)]{Bregman:1966:PhD})
\begin{equation}
D_\Psi(x,y):=\Psi(x)-\Psi(y)-\sum\nolimits_{i=1}^n(x_i-y_i)(\grad\Psi(y_i)),
\label{eqn.Vainberg.Bregman.D.Psi}
\end{equation}
for $x,y\in\RR^n$ (or $x,y\in M$) and $d=n$, another going back to the Brunk--Ewing--Utz \cite[Eqn. (4.4)]{Brunk:Ewing:Utz:1957}
\begin{equation}
D_\Psi^\mu(x,y):=\int\nolimits_{\X}\mu(\xx)D_\Psi(x(\xx),y(\xx)),
\end{equation}
for $x,y:\X\ra\RR$, $n=1$, $d=m$, and a measure $\mu$ on the Borel subsets of $\X\subseteq\RR^m$. The former approach has been generalised and widely developed for $\RR^n$ replaced by a reflexive Banach space $(X,\n{\cdot}_X)$ (see Section \ref{section.bregman.reflexive} and references therein). The latter approach has been generalised and further developed for $(\X,\mu)$ given by a countably finite nonzero measure space (see \cite{Csiszar:1995,Csiszar:Matus:2008,Csiszar:Matus:2009,Csiszar:Matus:2012:kybernetika,Stummer:Vajda:2012,Csiszar:Matus:2016}). These two approaches use different tools and have different assumptions and results: the former is developed within the frames of convex analysis on Banach spaces, and its vast range of results\footnote{One of the main virtues of the reflexive Banach space approach to $D_\Psi$ is that it generalises the theory of metric projections onto convex and closed subsets of a Hilbert space to the theory of (right and left) $D_\Psi$-projections on a reflexive Banach space $(X,\n{\cdot}_X)$, cf. \cite[\S\S4--6, \S\S7--8]{Alber:1993}. Furthermore, if $W\subseteq X$, then a Gateaux differentiable function $\Psi$ on $X$ determines a weakened notion of orthogonality, that can be seen as a weakening of the Hilbert space scalar product to the Banach space duality `shifted' by the Gateaux derivative of $\Psi$ (i.e. $\duality{\,\cdot\,,\DG\Psi(\cdot)}_{X\times X^\star}$), cf. \cite[Thm. 3.19, Rem. 3.20]{Alber:2007}. The (asymmetric and nonlinear) analogues of cosine and pythagorean theorems for $D_\Psi$-projections can be seen as a consequence of a definition of $D_\Psi$ via first order Taylor expansion of $\Psi$, combined with this notion of orthogonality.} relies heavily on topological and geometric properties exhibited by the reflexive $(X,\n{\cdot}_X)$, while the latter is measure theoretic, and depends on the properties of integrands of the convex functions, $\int\mu\Psi(\cdot):L_1(\X,\mu)\ra\,]-\infty,\infty]$ (which is a troublesome subject, cf., e.g., \cite{Rockafellar:1968,Teboulle:Vajda:1993,Borwein:Limber:1996,Borwein:Yao:2014}), with $(L_1(\X,\mu),\n{\cdot}_1:=\int\mu\ab{\cdot})$ being nonreflexive. 

The passage from probabilistic (measure theoretic) to quantum (operator algebraic) setting corresponds to replacing $(L_1(\X,\mu),\n{\cdot}_1)$ by the Banach predual $(\N_\star,\n{\cdot}_1:=\n{\cdot}_{\N_\star})$ of a W$^*$-algebra $(\N,\n{\cdot}_\N)$. Except of the finite dimensional cases, all of these spaces are nonreflexive. The noncommutative analogue $D^{\tr_\H}_\Psi$ of $D_\Psi^\mu$ was  introduced in \cite[\S2.2]{Tsuda:Raetsch:Warmuth:2005} for symmetric positive definite matrices, and in \cite[pp. 127--129]{Petz:2007}\footnote{More precisely, this paper introduces $D_\Psi^{\tr_\H,+}(x,y):=\tr_\H(D^+_\Psi(x,y))$ for a convex $\Psi:U\ra\BH$, where $U$ is a convex subset of a Banach space, e.g. $U=((\BH)_\star)^+$, $\BH$ is the W$^*$-algebra of bounded operators on a Hilbert space $\H$, and $D_\Psi^+$ is given by \eqref{eqn.right.Gateaux}. The evaluation of $D_\Psi^{\tr_\H,+}(x,y)$ is thus defined by spectral calculus applied to $\Psi$. This functional has been further investigated in \cite{Lewin:Sabin:2014,Pitrik:Virosztek:2015}.} for preduals of type I W$^*$-algebras (see also \cite[\S{}V]{Harremoes:2017:IEEE} (resp., \cite[\S\S8--9]{Harremoes:2018}) for preduals of type I$_n$ (resp., I$_2$) JBW-algebras with $n\in\NN$). However, due to nonreflexivity of $(\N_\star,\n{\cdot}_1)$, this definition shares the same convex theoretic problems as $D^\mu_\Psi$, and is incapable of utilising the results obtained for $D_\Psi$ over reflexive Banach spaces. It is also unclear how to extend the definition of $D_\Psi^{\tr_\H}$ to arbitrary W$^*$-algebras.

\subsection{Va\u{\i}nberg--Br\`{e}gman functionals over dually flat manifolds}

Let $M$ be a $\mathrm{C}^3$-manifold with a tangent bundle $\T M$, a $\mathrm{C}^3$ riemannian metric tensor $\gbold$ on $\T M$, and a pair $(\nabla,\widetilde{\nabla})$ of $\mathrm{C}^3$ affine connections on $\T M$ (with arbitrary curvatures and torsions). Let $\transport^\nabla_c$ denote a $\nabla$-parallel transport in $\T M$ along a curve $c$ in $M$. Then the \df{Norden--Sen geometry} is defined as a quadruple $(M,\gbold,\nabla,\widetilde{\nabla})$ satisfying any of the equivalent conditions \cite[pp. 205--206, \S2, \S4]{Norden:1937} \cite[p. 46]{Sen:1948}:\footnote{In comparison, given $(M,\gbold)$, the Levi-Civita affine connection $\nabla^\gbold$ \cite[Eqn. (A)]{LeviCivita:1917} is characterised among all torsion-free affine connections on $\T M$ by $\gbold(\transport^{\nabla^\gbold}_c(\cdot),\transport^{\nabla^\gbold}_c(\cdot))=\gbold$. Each torsion-free Norden--Sen geometry determines $\nabla^{\gbold}$ by $\nabla^{\gbold}=\frac{1}{2}(\nabla+\widetilde{\nabla})$ \cite[p. 211]{Norden:1937}.}
\begin{align}
\gbold(\transport^{\nabla}_c(\cdot),\transport^{\widetilde{\nabla}}_c(\cdot))&=\gbold,\\
\gbold(\nabla_uv,w)+\gbold(v,\widetilde{\nabla}_uw)&=u(\gbold(v,w))\;\forall u,v,w\in\T M.
\label{eqn.norden.sen}
\end{align}

If $Z$ is a finite dimensional $\mathrm{C}^3$-manifold and an information $D\in\mathrm{C}^3(Z\times Z;\RR^+)$ has a positive definite hessian matrix, then a third order Taylor expansion of $D$ on $Z$ induces \cite[pp. 795--796]{Eguchi:1983} a riemannian metric $\gbold^D$ on $\T Z$ and a pair $(\nabla^D,\widetilde{\nabla}^D)$ of torsion-free affine connections on $\T Z$, satisfying the characteristic property \eqref{eqn.norden.sen} of the Norden--Sen geometry. This way the global geometric properties of $D$ can be analysed in local terms of its torsion-free Norden--Sen differential geometry.\footnote{Following \cite[\S4]{Lauritzen:1987:statistical:manifolds}, the torsion-free Norden--Sen geometries are sometimes called ``statistical manifolds''. Apart from not crediting the original authors of this structure, such terminology is misleading, since these geometries are independent of any notion of statistics.} 

The \df{dually flat} (a.k.a. \df{hessian}) geometry \cite[Prop. (p. 213)]{Shima:1976} is characterised among all torsion-free Norden--Sen geometries by the flatness of $\nabla$ and $\widetilde{\nabla}$. This is equivalent \cite[\S\S3.2--3.3]{Nagaoka:Amari:1982} with existence of two coordinate systems, $\{\theta_i\mid i\in\{1,\ldots,n\}\}:M\ra\RR^n$ and $\{\eta_i\mid i\in\{1,\ldots,n\}\}:M\ra\RR^n$, such that, $\forall\rho\in M$, 
\begin{empheq}[left={ \empheqlbrace}]{align}
\eta_i(\rho)&=\frac{\partial\Psi(\theta(\rho))}{\partial\theta^i}=((\grad\Psi\circ\theta)(\rho))_i\label{eqn.eta.Phi}\\
\theta_i(\rho)&=\frac{\partial\Psi^\lfdual(\eta(\rho))}{\partial\eta^i}=((\grad\Psi^\lfdual\circ\eta)(\rho))_i\label{eqn.theta.Phi}\\
\Psi^\lfdual(y)&=\sup_{x\in\RR^n}\left\{\sum_{i=1}^nx_iy_i-\Psi(x)\right\}\; \forall y\in\RR^n,\label{eqn.finite.Fenchel.duality}
\end{empheq}
and, for $D_{\theta,\Psi}(\rho,\sigma):=D_\Psi(\theta(\rho),\theta(\sigma))$ with $D_\Psi$ defined by \eqref{eqn.Vainberg.Bregman.D.Psi},  
\begin{empheq}[left={ \empheqlbrace}]{align}
\Upgamma^{\nabla^{D_{\theta,\Psi}}}_{ijk}(\theta(\rho))&=0,\;\;\Upgamma^{\widetilde{\nabla}^{D_{\theta,\Psi}}}_{ijk}(\eta(\rho))=0\label{eqn.upgamma.nabla}\\
\gbold^{D_{\theta,\Psi}}_{ij}(\theta(\rho))&=\frac{\partial^2\Psi(\theta(\rho))}{\partial\theta^i\partial\theta^j},\label{eqn.gbold.D.theta.Psi}
\end{empheq}
where $\Upgamma^\nabla(u,v,w):=\gbold(\nabla_uv,w)$ $\forall u,v,w\in\T M$, while the subscript ${}_i$ denotes evaluation at the $i$-th component of a basis in $\T M$ given by coordinate system differentials (i.e. setting $u=\frac{\partial}{\partial\theta^i}$, $v=\frac{\partial}{\partial\theta^j}$, etc., for $\theta(\rho)$, and analogously for $\eta(\rho)$).\footnote{Due to \eqref{eqn.eta.Phi}, \eqref{eqn.theta.Phi}, and \eqref{eqn.finite.Fenchel.duality}, \eqref{eqn.gbold.D.theta.Psi} is equivalent with $\gbold^{D_{\eta,\Psi^\lfdual}}_{ij}(\eta(\rho))=\frac{\partial^2\Psi^\lfdual(\eta(\rho))}{\partial\eta^i\partial\eta^j}$, where $D_{\theta,\Psi}=D_{\eta,\Psi^\lfdual}((\grad\Psi)(\cdot),(\grad\Psi)(\cdot))$.} When considered in this setting, the left (resp., right) generalised py\-tha\-go\-re\-an equation \eqref{eqn.gen.pyth.left} (resp., \eqref{eqn.gen.pyth.right}) implies a corresponding equation for $D_{\theta,\Psi}$, and the latter is equivalent with \cite[Cor. 3.5]{Amari:Nagaoka:1993}: a projection of $\rho\in M$ onto $\nabla^{D_{\theta,\Psi}}$-(resp., $\widetilde{\nabla}^{D_{\theta,\Psi}}$-)autoparallel submanifold $C$ along $\widetilde{\nabla}^{D_{\theta,\Psi}}$-(resp., $\nabla^{D_{\theta,\Psi}}$-)geodesics is $\gbold^{D_{\theta,\Psi}}$-orthogonal (= $\gbold^{D_{\eta,\Psi^\lfdual}}$-orthogonal) to $C$. 

\subsection{Our approach}

In Section \ref{section.bregman.nonreflexive} we present a new approach to extension of $D_\Psi$ to nonreflexive Banach spaces $(Y,\n{\cdot}_Y)$, by means of nonlinear embedding $\ell:Z\ra X$, where $Z\subseteq Y$ and $(X,\n{\cdot}_X)$ is a reflexive Banach space. The main idea is to pull back the properties exhibited by $D_\Psi$ on $(X,\n{\cdot}_X)$ into the corresponding properties exhibited by $D_{\ell,\Psi}(\cdot,\cdot):=D_\Psi(\ell(\cdot),\ell(\cdot))$ on $Z$. In general, $\ell$ act as (global or local) coordinate systems on the subsets of $(Y,\n{\cdot}_Y)$, while $(X,\n{\cdot}_X)$ are (global or local) codomains of these systems. As a result, we deal with \textit{$\ell$-convex} (resp., \textit{$\ell$-closed}) sets in $(Y,\n{\cdot}_Y)$, i.e. the preimages of convex (resp., closed) sets in $(X,\n{\cdot}_X)$ with respect to $\ell$.\footnote{As already featured in \eqref{eqn.gen.pyth.right}, consideration of right $D_\Psi$-projections introduces also $\DG\Psi$-closed and $\DG\Psi$-convex subsets of $(X,\n{\cdot}_X)$, giving rise to $\DG\Psi\circ\ell$-closed and $\DG\Psi\circ\ell$-convex subsets of $Z$.} 

In order to express topological behaviour of $D_{\ell,\Psi}$ in terms of $(Y,\n{\cdot}_Y)$, without relativisation to $(X,\n{\cdot}_X)$, $\ell$ and its inverse, $\ell^\inver$, have to additionally preserve the corresponding continuity properties. In such case the $\ell$-closed sets coincide with the sets in $Z$ that are closed with respect to the topology of $\n{\cdot}_Y$. Hence, the best behaved sector of the theory of extended Va\u{\i}nberg--Br\`{e}gman information $D_{\ell,\Psi}$ belongs to an intersection of nonlinear convex analysis on reflexive Banach spaces with a nonlinear homeomorphic theory of Banach spaces. (E.g., if $\ell$ and $\ell^\inver$ exhibit Lipschitz--H\"{o}lder continuity, then the results on Lipschitz--H\"{o}lder continuity of $\LPPP^{D_{\Psi}}_K$ and $\RPPP^{D_{\Psi}}_K$ can be pulled back into the corresponding results on Lipschitz--H\"{o}lder continuity of $\LPPP^{D_{\ell,\Psi}}_{\ell^\inver(K)}$ and $\RPPP^{D_{\ell,\Psi}}_{\ell^\inver(K)}$.) 

On the other hand, the relativisation of convexity to $\ell$-convex sets is unavoidable. One can interpret $\ell$-convex sets as an optimal data type for statistical inference (e.g., for the setting based on $D_\Psi$-projections) that is carried on $(Y,\n{\cdot}_Y)$ through the ``lenses'' of $\ell$. For a given pair of $(Y,\n{\cdot}_Y)$ and $(X,\n{\cdot}_X)$ there may be more than one map $\ell$ that is relevant for consideration. For example (see Proposition \ref{prop.when.lozanovskii.equals.kaczmarz}), for $(X,\n{\cdot}_X)$ given by (commutative or noncommutative) Orlicz spaces, the Kaczmarz map generally differs from the Lozanovski\u{\i} factorisation map (except of the case of $L_p$ spaces, when both of these maps coincide with the Mazur map), while both exhibit uniform homeomorphy on unit spheres under suitable conditions. 

For $X=\RR^n$, $D_{\ell,\Psi}$ recovers the setting of Va\u{\i}nberg--Br\`{e}gman information $D_{\theta,\Psi}$ on an $n$-di\-men\-sio\-nal $\mathrm{C}^1$-manifold $M$, with the map $\ell:M\ra\RR^n$ (resp., $\DG\Psi\circ\ell:M\ra\RR^n$) given by the coordinate system $\{\theta_i\}$ (resp., $\{\eta_i\}$). More specifically, a domain $M$ of a dually flat geometry is assumed to be a (suitably differentiable) manifold, covered by two globally affine coordinates $\{\theta_i\}$ and $\{\eta_i\}$, \textit{without} assuming $M\subseteq\RR^n$, cf. \cite[p. 213]{Shima:1976} and \cite[\S3.2]{Amari:Nagaoka:1993}. This is not addressed by \eqref{eqn.Vainberg.Bregman.D.Psi}, and is addressed (up to a weaker assumption on the order of differentiability) by \eqref{eqn.generalised.Bregman}. Equation \eqref{eqn.finite.Fenchel.duality} is a special case of \eqref{eqn.Fenchel.duality}. Furthermore, gradients in \eqref{eqn.eta.Phi} and \eqref{eqn.theta.Phi} require only $\mathrm{C}^1$-differentiability of $\Psi$ and $\Psi^\lfdual$. The approach of Section \ref{section.bregman.nonreflexive} is rooted in an observation that the Banach space generalisation of \eqref{eqn.eta.Phi} and \eqref{eqn.theta.Phi} requires two components: Euler--Legendre $\Psi$ on a reflexive Banach space $(X,\n{\cdot}_X)$, and nonlinear embeddings into $(X,\n{\cdot}_X)$ and $(X^\star,\n{\cdot}_{X^\star})$, given by $\ell$ and $\DG\Psi\circ\ell$, respectively, and replacing, respectively, $\theta$ and $\eta$. This way the framework of extended Va\u{\i}nberg--Br\`{e}gman information $D_{\ell,\Psi}$ unifies reflexive Banach space theoretic and (the convex/$\mathrm{C}^1$ layer of) finite dimensional smooth information geometric approaches to Va\u{\i}nberg--Br\`{e}gman information.

Looking at the origins of the finite dimensional commutative information geometry (so, with an atomic finite $\X$ and counting measure $\mu$), $\ell$ can be traced back to identification of Fisher's `log-likelihood' $\log(p)$ \cite[\S3]{Fisher:1912} as a coordinate system on a manifold $W\subseteq(S(L_1(\X,\mu),\n{\cdot}_1))^+$ of probability densities, cf., e.g., \cite[\S10.3]{Chencov:1972}. For $\ell(p)=\log(p)$, $W$ given by an exponential family (of probability densities) is a totally geodesic surface\footnote{For any torsion-free affine connection $\nabla$ on a $\mathrm{C}^3$ manifold $M$, the submanifolds of $M$ that are totally geodesic with respect to $\nabla$ coincide with the $\nabla$-autoparallel submanifolds of $M$.} of the affine connection $\nabla^0$ \cite[p. 544]{Chencov:1964} (which arises, e.g., as $\nabla^{D_1}$, where $D_1$ is the Kullback--Leibler information), while for $\ell(p)=\ell_\gamma(p)=p^\gamma$, the $\ell$-affine subsets are given by \cite[Thm. 6.3]{Morozova:Chencov:1991} the totally geodesic surfaces of the corresponding Chencov--Amari affine connections $\nabla^\gamma$.\footnote{$\nabla^\gamma$, with $\gamma\in\RR$, were introduced and characterised in \cite[Thm. 12.2]{Chencov:1972}, and later independently rediscovered in \cite[Def. 2]{Amari:1980}. $\nabla^\gamma=\nabla^{D_\gamma}$, where $D_\gamma$ is given by the finite dimensional commutative version of \eqref{eqn.d.gamma} (up to an extension from $\gamma\in\,]0,1[$ to $\gamma\in\RR$, which is possible due to finite dimensionality). In the finite dimensional  noncommutative case for type I$_n$ W$^*$-algebras, the analogues of $\nabla^1$ and $\nabla^\gamma$ were introduced, respectively, in \cite[p. 450]{Nagaoka:1995} \cite[\S7]{Hasegawa:1995} and \cite[\S5]{Hasegawa:1995}, and are equal, respectively, to $\nabla^{D_1}$ and $\nabla^{D_\gamma}$, where $D_1$ is the Umegaki information \cite[Def. 1]{Umegaki:1961} (i.e. $D_1(\rho,\sigma):=\tr_\H(\rho\log\rho-\rho\log\sigma)$ $\forall\rho,\sigma\in(S((\BH)_\star,\n{\cdot}_1))^+$).} Hence, the $\ell$-affine (resp., $\DG\Psi\circ\ell$-affine) subsets of $(V,\n{\cdot}_V)$ can be seen as a direct generalisation of the exponential (resp., $\widetilde{\nabla}^{D_1}$-autoparallel) families (in particular), and of the $\nabla$-autoparallel (resp., $\widetilde{\nabla}$-autoparallel) torsion-free Norden--Sen information manifolds (in general). Further passage to $\ell$-convex (resp., $\DG\Psi\circ\ell$-convex) subsets generalises $\nabla$(resp., $\widetilde{\nabla}$)-geodesically convex torsion-free Norden--Sen information manifolds, and can be seen as a natural step that takes into account the convex theoretic setting of various statistical inference problems.

In this context, our approach can be seen as partially arising from an observation that the $\ell_\gamma$ (resp., $\ell_\orlicz$) embeddings, used in \cite[Eqn. (2.7)]{Nagaoka:Amari:1982} (resp., \cite[\S7.2]{Gibilisco:Pistone:1998}), are Mazur (resp., Kaczmarz) maps \cite[p. 83]{Mazur:1929} (resp., \cite[p. 148]{Kaczmarz:1933}) on $(L_1(\X,\mu))^+$. In particular, the abstract framework proposed by us in \cite[Eqns. (24), (31)]{Kostecki:2011:OSID} has resulted from rethinking of an important example in \cite[\S6--\S8]{Jencova:2005} (see Remark \ref{remark.relation.with.jencova.results}). The current paper, through Definition \ref{def.ell.Psi.information} and Proposition \ref{prop.D.ell.psi.properties}, provides a concrete functional analytic implementation of this abstract framework, based on the use of Euler--Legendre $\Psi$ and totally convex $\Psi^\lfdual$.\footnote{More precisely, restricting the discussion in \cite[\S3]{Kostecki:2011:OSID} to $L=(X,\n{\cdot}_X)$ and $\duality{\cdot,\cdot}_L=\duality{\cdot,\cdot}_{X\times X^\star}$, with a reflexive Banach space $(X,\n{\cdot}_X)$, setting a dualiser $f_\Psi$ to be given by $\DG\Psi$ for $\Psi\in\pclg(X,\n{\cdot}_X)$, setting a dual embedding $\ell^@$ to be given by $\DG\Psi\circ\ell$, and requiring $\Psi$ to be Euler--Legendre, gives the basic setting of the current paper.}


In order to deliver concrete examples for various spaces $(X,\n{\cdot}_X)$, we consider $\Psi=\Psi_{\alpha,\beta}:=\frac{\beta}{\alpha}\n{\cdot}_X^{1/\beta}$ with $(\alpha,\beta)\in\,]0,\infty[\,\times\,]0,1[$. As discussed in Section \ref{section.bregman.reflexive}, the key benefit of $\Psi=\Psi_{\alpha,\beta}$ is that it boils down the task of establishing that $\Psi$ is Euler--Legendre and $\Psi^\lfdual$ is totally convex (implying the fine behaviour of $D_\Psi$) into the task of establishing corresponding geometric properties of $(X,\n{\cdot}_X)$. As a result, the main body of proofs carried in Section \ref{section.nonreflexive.applications} amounts to using the geometric properties of the corresponding Banach spaces $(X,\n{\cdot}_X)$.

In Section \ref{section.nonreflexive.applications} we apply this approach to derive some new families of extended Va\u{\i}nberg--Br\`{e}gman relative entropies over some radially compact base normed spaces $(V,\n{\cdot}_V)$ in spectral duality, with state spaces given by bases $(S(V,\n{\cdot}_V))^+:=\{\phi\in V^+\mid\n{\phi}_V=1\}$. In general, with an exception of the finite dimensional case, $(V,\n{\cdot}_V)$ are not reflexive. We deal with $Z\subseteq V$ whenever possible; the restriction to $Z\subseteq V^+$ or $Z\subseteq(S(V,\n{\cdot}_V))^+$ is straightforward.

We start from consideration of three special cases of $(V,\n{\cdot}_V)$:  the preduals of any W$^*$-algebras and of semifinite JBW-algebras and the preduals of generalised spin factors $(X^\star\oplus\RR,\n{\cdot}_{X^\star\oplus\RR})$. 

In the first two cases, we first implement $\ell$ by the corresponding generalisations $\ell_\gamma$ of a Mazur map \cite[p. 83]{Mazur:1929} into $L_p$ spaces over these algebras, with $1/\gamma:=p\in\,]1,\infty[$. For semifinite W$^*$-algebras $\N$ we consider also $\ell$ given by the noncommutative generalisation $\ell_\orlicz$ of a Kaczmarz map \cite[p. 148]{Kaczmarz:1933} into noncommutative Orlicz spaces over $\N$. The noncommutative Kaczmarz maps and the nonassociative Mazur maps have not been considered before. We prove the Lipschitz--H\"{o}lder continuity of the latter on the positive parts of unit balls.\footnote{In commutative (resp., noncommutative) case, the Lipschitz--H\"{o}lder continuity of Mazur maps on unit balls was established in \cite[p. 449]{Stone:1948:II} (resp., \cite[Thm. (p. 37)]{Ricard:2015}).} Furthermore, we characterise strict convexity, Gateaux differentiability, Radon--Riesz--Shmul'yan property, and reflexivity of Orlicz and Morse--Transue--Nakano norms on noncommutative Orlicz spaces. 

In the latter case, $\ell$ is implemented by $\ell_{/\RR}(x):=(x,1)$. Except of the case when $(X,\n{\cdot}_X)$ is a Hilbert space (see Remark \ref{remark.spin.factors.are.JBW.algebras}), the generalised spin factors $(X^\star\oplus\RR,\n{\cdot}_{X^\star\oplus\RR})$ are not JBW-algebras, so they provide an example of a `postquantum theory' beyond JBW-algebras. In Remark \ref{remark.postquantum.Lp}, Proposition \ref{prop.berdikulov.A.gamma}, and Definition \ref{def.L.gamma.order.unit} we take first steps towards more systematic treatment of postquantum models of $D_{\ell,\Psi}$ based on a corresponding generalisation of the Mazur map. 

Furthermore, in order to show that the framework of $D_{\ell,\Psi}$ is not limited to $\ell$ given by $\ell_\gamma$, $\ell_\orlicz$, or $\ell_{/\RR}$ (and, thus, either to quite special spaces $(X,\n{\cdot}_X)$ or to quite special spaces $(V,\n{\cdot}_V)$), we consider also a case when $(V,\n{\cdot}_V)=(L_1(\X,\mu),\n{\cdot}_1)$ (resp., $(V,\n{\cdot}_V)=(\BH_\star,\n{\cdot}_1)$) for a localisable measure space $(\X,\mu)$ (resp., for a Hilbert space $\H$ with $\dim\H\in\NN$), while $\ell$ is given by the Lozanovski\u{\i} factorisation map $\ell_E$ \cite[Thm. 6.3]{Lozanovskii:1969} \cite[\S3.(a)]{Gillespie:1981} (resp., \cite[Prop. 5.5]{Dodds:Dodds:dePagter:1992}) into an arbitrary Banach function space $(E(\X,\mu),\n{\cdot}_{E(\X,\mu)})$ (resp., an arbitrary unitarily invariant ideal of compact operators over $\H$) that is uniformly convex and uniformly Fr\'{e}chet differentiable. This is the first appearance of the Lozanovski\u{\i} factorisation map in information theory. However, quite strikingly, the explicit formula for this map, \eqref{eqn.lozanovskii.ell.RPPP.D.1} in commutative case and \eqref{eqn.nc.lozanovskii.ell.RPPP.D.1} in noncommutative case, is equal to a right entropic projection for a variant of, respectively, the Kullback--Leibler information and the Umegaki information.\footnote{While Chencov introduced $\RPPP^{D_1}_K$ in 1968 \cite[p. 328]{Chencov:1968} and Lozanovski\u{\i} introduced $\ell_E$ in 1969 \cite[Thm. 6.3]{Lozanovskii:1969}, the explicit identification of the latter map as a suitable variant of the former one has appeared only in \cite[\S5]{ChavezDominguez:2023}, over a half of a century later. (Strictly speaking, \cite[Thm. 6.3]{Lozanovskii:1969} does not contain the statement of uniqueness of factorisation. The uniqueness was stated without a proof in \cite[1117]{Lozanovskii:1969:VIII} and was proved in \cite[\S3.(a)]{Gillespie:1981}.)} As a result, the left and right $D_{\ell_E,\Psi}$-projections, considered in Propositions \ref{prop.lozanovskii.D.ell.Psi} and \ref{prop.nc.lozanovskii.D.ell.Psi}, have very interesting explicit form: $\LPPP_C^{D_{\RPPP^{\bar{D}_1}_K,\Psi}}$ and $\RPPP_C^{D_{\RPPP^{\bar{D}_1}_K,\Psi}}$, respectively. 

\section{Extended Va\u{\i}nberg--Br\`{e}gman functionals}\label{section.two.approaches.vainberg.bregman}

\subsection{Banach space theoretic preliminaries}

In what follows, $(X,\n{\cdot}_X)$ will denote a Banach space over $\KK\in\{\RR,\CC\}$ \cite[\S1]{Banach:1922}, $B(X,\n{\cdot}_X):=\{x\in X\mid \n{x}_X\leq1\}$, $S(X,\n{\cdot}_X):=\{x\in X\mid\n{x}_X=1\}$. $(X^\star,\n{\cdot}_{X^\star})$ will denote a Banach space of continuous linear functions $X\ra\KK$, equipped with a norm $\n{y}_{X^\star}:=\sup\{\ab{y(x)}\mid x\in B(X,\n{\cdot}_X)\}$ $\forall y\in X^\star$ \cite[p. 62]{Helly:1921}, and will be called a \df{Banach dual} of $(X,\n{\cdot}_X)$ (with respect to a bilinear duality $\duality{x,y}_{X\times X^\star}:=y(x)\in\KK$ $\forall x\in X$ $\forall y\in X^\star$). If $(Y,\n{\cdot}_Y)$ and $(X,\n{\cdot}_X)$ are such that $(Y^\star,\n{\cdot}_{Y^\star})=(X,\n{\cdot}_X)$, then $Y=:X_\star$ is called a \df{predual} of $X$. $(X,\n{\cdot}_X)$ is said to be \df{reflexive} if{}f \cite[pp. 219--220]{Hahn:1927} $X\ni x\mapsto\duality{x,\,\cdot\,}_{X\times X^\star}\in X^\star{}^\star$ is an isometric isomorphism. Given $(X,\n{\cdot}_X)$, for any $Y\subseteq X$, $\INT(Y)$ (resp., $\overline{Y}^{\n{\cdot}_X}$) will denote a topological interior (resp., a topological closure) of $Y$ with respect to the topology of $\n{\cdot}_X$. 

A Banach space $(X,\n{\cdot}_X)$ is said to satisfy the \df{Radon--Riesz--Shmul'yan property}\footnote{This property was first considered by Radon \cite[p. 1363]{Radon:1913} and Riesz \cite[p. 182]{Riesz:1929} for commutative $L_{1/\gamma}$ spaces, $\gamma\in\,]1,\infty[$. For the general Banach spaces it was first introduced and studied by Shmul'yan in \cite[Thm. 5]{Shmulyan:1939:geometrical}. Sometimes it is called a ``Kadec--Klee property'', however Kadec \cite[p. 13]{Kadec:1958} explicitly refers to this work of Shmul'yan, while Klee \cite[pp. 25--27]{Klee:1960} explicitly refers to this paper by Kadec.} if{}f \cite[Thm. 5]{Shmulyan:1939:geometrical}, for any $\{x_n\in X\mid n\in\NN\}$, convergence of $x_n$ to $x\in X$ in a weak topology together with $\lim_{n\ra\infty}\n{x_n}_X=\n{x}_X$ implies $\lim_{n\ra\infty}\n{x_n-x}_X=0$. A Banach space $(X,\n{\cdot}_X)$ is called: \df{strictly convex} if{}f \cite[p. 39]{Frechet:1925:affines} $\forall x,y\in X\setminus\{0\}$
\begin{equation}
\n{x+y}_X=\n{x}_X+\n{y}_X\;\limp\;\exists\lambda>0\;y=\lambda x;
\end{equation}
\df{uniformly convex} if{}f \cite[Def. 1]{Clarkson:1936} $\forall\epsilon_1>0$
\begin{equation}
\exists\epsilon_2>0\;\forall x,y\in S(X,\n{\cdot}_X)\;\n{x-y}_X\geq\epsilon_1\;\limp\;\textstyle\frac{1}{2}\n{x+y}_X\leq1-\epsilon_2;
\end{equation}
\df{uniformly Fr\'{e}chet differentiable} if{}f \cite[p. 375]{Day:1944} $\forall\epsilon_1>0$
\begin{equation}
\exists\epsilon_2>0\;\forall x,y\in S(X,\n{\cdot}_X)\;\n{x-y}_X\leq\epsilon_1\limp1-\textstyle\frac{1}{2}\n{x+y}_X\leq\epsilon_2\n{x-y}_X.\!\!
\end{equation}
Given Banach spaces $(X,\n{\cdot}_X)$ and $(Y,\n{\cdot}_Y)$, $Z\subseteq X$, $W\subseteq Y$, $t\in\,]0,\infty[$, a function $f:Z\ra W$ is said to be \df{$t$-Lipschitz--H\"{o}lder continuous} on $Z$ if{}f $ \exists c>0$ $\forall x,y\in Z$ $\n{f(x)-f(y)}_Y\leq c\n{x-y}_X^t$ (if $t=1$ then $f$ is said to be \df{Lipschitz continuous}).\footnote{This condition, both for $t=1$ and $t<1$, has been introduced by Lipschitz in \cite[Eqn. (2)]{Lipschitz:1876} and \cite[Eqn. (2$^\star$)]{Lipschitz:1876}, respectively, six years before H\"{o}lder's \cite[pp. 17--18]{Hoelder:1882}.} For convenience of notation, in what follows we assume that $\KK=\RR$. We also assume $\inf\varnothing:=\infty$.

Given a Banach space $(X,\n{\cdot}_X)$, $\Psi:X\ra\,]-\infty,\infty]$ will be called: \df{proper} if{}f $\efd(\Psi):=\{x\in X\mid\Psi(x)\neq\infty\}\neq\varnothing$; \df{convex} (resp., \df{strictly convex}) if{}f  $x\neq y$ $\limp$ $\Psi(\lambda x+(1-\lambda)y)\leq\mbox{ (resp., }<\mbox{) }\lambda\Psi(x)+(1-\lambda)\Psi(y)$ $\forall x,y\in\efd(\Psi)$ $\forall\lambda\in\,]0,1[$.\footnote{This is equivalent to the definition based on the same inequality, with quantifiers changed to $\forall x,y\in X$ $\forall\lambda\in[0,1]$, with the conventions $\infty+\infty\equiv\infty$, $0\cdot\infty\equiv\infty$, $0\cdot(-\infty)\equiv0$, and without assuming $x\neq y$.} The set of all proper, convex, lower semicontinuous (with respect to $\n{\cdot}_X$) functions $\Psi:X\ra\,]-\infty,\infty]$ will be denoted by $\pcl(X,\n{\cdot}_X)$. If $\Psi:X\ra\,]-\infty,\infty]$ is proper, then the \df{right Gateaux derivative} of $\Psi$ at $x\in\efd(\Psi)$ in the direction $h\in X$ reads \cite[p. 53]{Ascoli:1932} $\forall(x,h)\in\efd(\Psi)\times X$
\begin{equation}
\DG_+\Psi(x,h):=\lim_{t\ra^+0}\textstyle\frac{1}{t}(\Psi(x+th)-\Psi(x))\in\,]-\infty,\infty],
\end{equation}
and it exists $\forall h\in X$. $\Psi\in\pcl(X,\n{\cdot}_X)$ is called \df{Gateaux differentiable} at $x\in\intefd{\Psi}$ if{}f \cite[p. 311]{Gateaux:1914} $\DG_+\Psi(x,y)=-\DG_+\Psi(x,-y)$ $\forall y\in X$. In such case $\DG_+\Psi(x,\,\cdot\,)$ is linear, so it defines a bounded linear operator $\DG_+\Psi(x,y)=:\duality{y,\DG\Psi(x)}_{X\times X^\star}$ $\forall y\in X$. A set of all $\Psi\in\pcl(X,\n{\cdot}_X)$ such that are Gateaux differentiable on $\intefd{\Psi}\neq\varnothing$ will be denoted $\pclg(X,\n{\cdot}_X)$. A Banach space $(X,\n{\cdot}_X)$ is called: \df{Gateaux differentiable} if{}f \cite[p. 78]{Mazur:1933} $\n{\cdot}_X$ is Gateaux differentiable at every $x\in X\setminus\{0\}$; \df{Fr\'{e}chet differentiable} if{}f \cite[p. 129]{Mazur:1933:schwache}, for any fixed $x\in X\setminus\{0\}$, $\DG\n{h}_X(x)$ exist in uniform convergence $\forall h\in S(X,\n{\cdot}_X)$. In the latter case $\DG\n{\cdot}_X$ will be denoted by $\DF\n{\cdot}_X$.

For a proper $\Psi:X\ra\,]-\infty,\infty]$,
\begin{equation}
X^\star\ni y\mapsto\Psi^\lfdual(y):=\sup_{x\in X}\{\duality{x,y}_{X\times X^\star}-\Psi(x)\}\in\,]-\infty,\infty],
\label{eqn.Fenchel.duality}
\end{equation}
called a \df{Fenchel dual} of $\Psi$ \cite[p. 75]{Fenchel:1949} \cite[p. 8]{Moreau:1962} (cf. \textup{\cite[Eqn. (1)]{Mandelbrojt:1939}}), satisfies $\Psi^\lfdual\in\pcl(X^\star,\n{\cdot}_{X^\star})$ \cite[Thm. 3.6]{Broendsted:1964}. If $(X,\n{\cdot}_X)$ is reflexive and $\Psi\in\pclg(X,\n{\cdot}_X)$, then $\Psi$ will be called \df{Euler--Legendre}\footnote{These functions are usually called ``Legendre'' (although they were introduced namelessly for $X=\RR^n$ in \cite[Thm. C-K]{Rockafellar:1963}). Yet, the transformation $\dd(z(x,y)-px-qy)=-x\dd p-y\dd q$, with $p=\frac{\partial z(x,y)}{\partial x}$ and $q=\frac{\partial z(x,y)}{\partial y}$, was introduced first by Euler \cite[Probl. 11 (Part I)]{Euler:1770}, and only 17 years later by Legendre \cite[p. 347]{Legendre:1787}.} if{}f \cite[Def. 5.2.(iii), Thms. 5.4, 5.6]{Bauschke:Borwein:Combettes:2001} \cite[\S2.1]{Reich:Sabach:2009} $\Psi^\lfdual\in\pclg(X^\star,\n{\cdot}_{X^\star})$, $\efd(\DG\Psi):=\{x\in\efd(\Psi)\mid\exists\;\DG\Psi(x)\}=\intefd{\Psi}$, and $\efd(\DG\Psi^\lfdual)=\intefd{\Psi^\lfdual}$. For $X=\RR^n$, this definition of Euler--Legendre functions goes back to Rockafellar, who showed \cite[Thm. C-K]{Rockafellar:1963} \cite[Thm. 1]{Rockafellar:1967} that if $\varnothing\neq U\subseteq\RR^n$ is open and convex, while $\Psi:U\ra\,]-\infty,\infty]$ is strictly convex, differentiable on $U$, and 
\begin{equation}
\lim_{t\ra^+0}\textstyle\frac{\dd}{\dd t}\Psi(tx+(1-t)y)=-\infty\;\forall(x,y)\in U\times(\overline{U}^{\n{\cdot}_{\RR^n}}\setminus U),
\end{equation}
then $\grad\,\Psi$ is a bijection on $U$, $\grad(\Psi^\lfdual)=(\grad\,\Psi)^\inver$ on $(\grad\,\Psi)(U)$, and $\Psi^\lfdual$ on $(\grad\,\Psi)(U)$ satisfies the same conditions as $\Psi$ on $U$. $\Psi\in\pcl(X,\n{\cdot}_X)$ is called \df{totally convex} at $x\in\efd(\Psi)$ if{}f  \cite[2.2]{Butnariu:Censor:Reich:1997} \cite[p. 62]{Butnariu:Iusem:1997} $\inf\{D^+_\Psi(y,x)\mid y\in\efd(\Psi),\;\n{y-x}_X=t\}>0$ $\forall t\in\,]0,\infty[$, where $D_\Psi^+:X\times X\ra[0,\infty]$ is given by \cite[Eqn. (2)]{Butnariu:Iusem:1997}
\begin{equation}
D_\Psi^+(x,y):=
\left\{\begin{array}{ll}
\Psi(x)-\Psi(y)-\DG_+\Psi(y;x-y)&
\st(x,y)\in X\times\efd(\Psi)\\
\infty&
\st\mbox{otherwise}.
\end{array}\right.
\label{eqn.right.Gateaux}
\end{equation}

\subsection{Va\u{\i}nberg--Br\`{e}gman functionals on reflexive Banach spaces}\label{section.bregman.reflexive}

\begin{definition}\label{def.Bregman.reflexive}
\textup{\cite[Eqn. (8.5)]{Vainberg:1956}}\footnote{\label{footnote.vainberg.bregman}This work contains already the Banach space formula \eqref{eqn.bregman}, that was further considered, e.g., in \cite[Lem. 1]{Vainberg:1965} and \cite[Thm. 3]{Kachurovskii:1966}. Ten years later, \cite[p. 1021]{Bregman:1966} (=\cite[Eqn. (2.1)]{Bregman:1966:PhD}) Br\`{e}gman independently introduced \eqref{eqn.Vainberg.Bregman.D.Psi}, that has received its Banach space formulation \eqref{eqn.bregman} in \cite[Eqn. (1)]{Alber:Butnariu:1997}, i.e. fourty years after Va\u{\i}nberg's \textup{\cite[Eqn. (8.5)]{Vainberg:1956}}.} For any $\Psi\in\pclg(X,\n{\cdot}_X)$ and any $x,y\in X$, the \df{Va\u{\i}nberg--Br\`{e}gman functional} on $(X,\n{\cdot}_X)$ is defined as $D_\Psi:X\times X\ra[0,\infty]$, where $\forall x\in X$ 
\begin{equation}
D_\Psi(x,y):=
        \left\{
                \begin{array}{ll}
                        \!\!\Psi(x)-\Psi(y)-\duality{x-y,\DG\Psi(y)}_{X\times X^\star}\!\!\!\! &\st y\in\intefd{\Psi}\\
                        \!\!\infty
												&\st\mbox{otherwise}.
                \end{array}
        \right.\!\!\!\!\!
\label{eqn.bregman}
\end{equation}
%
%
\end{definition}

\begin{proposition}\label{prop.D.psi.information}
\textup{\cite[Prop. 1.1.9]{Butnariu:Iusem:2000}}
If $\Psi\in\pclg(X,\n{\cdot}_X)$, then $D_\Psi$ is an information on $X$ if{}f $\Psi$ is strictly convex on $\intefd{\Psi}$.
\end{proposition}

\begin{definition}\label{def.d.projections}
Let $\Psi\in\pclg(X,\n{\cdot}_X)$, $y\in\intefd{\Psi}$, and $K\subseteq X$ with $\varnothing\neq K\cap\intefd{\Psi}$. If the set
$\arginff{x\in K\subseteq\intefd{\Psi}}{D_\Psi(y,x)}$
  (resp., $\arg$ $\inf_{x\in K}\{D_\Psi(x,y)\}$) is a singleton, then its element will be denoted $\RPPP^{D_\Psi}_K(y)$ (resp., $\LPPP^{D_\Psi}_K(y)$), and called a \df{right} (resp., \df{left}) \df{$D_\Psi$-projection} of $y$ onto $K$ \textup{\cite[Def. 3.1, Lem. 3.5]{Bauschke:Noll:2002}}\footnote{First special case of right $D_\Psi$-projection for nonsymmetric $D_\Psi$, with $D_\Psi$ given by the Kullback--Leibler information, was introduced in \cite[Eqn. (16)]{Chencov:1968} \cite[Def. 22.2]{Chencov:1972}.} (resp., \textup{\cite[p. 1019]{Bregman:1966} (=\cite[p. 14]{Bregman:1966:PhD})}\footnote{First special case of left $D_\Psi$-projection for nonsymmetric $D_\Psi$, with $D_\Psi$ given by the Kullback--Leibler information, was introduced in \cite[p. 32]{Sanov:1957} \cite[Ch. 3.2]{Kullback:1959}.}), while $K$ will be called a \df{right} (resp., \df{left}) \df{$D_\Psi$-Chebysh\"{e}v} set \textup{\cite[Def. 1.7]{Bauschke:Macklem:Wang:2011}} (resp., \textup{\cite[Def. 3.28]{Bauschke:Borwein:Combettes:2003}}).
\end{definition}

\begin{proposition}\label{prop.D.psi.projections}
\textup{\cite[Cor. 3.35]{Bauschke:Borwein:Combettes:2003}} 
If $(X,\n{\cdot}_X)$ is reflexive, $\Psi$ is Euler--Legendre, $\varnothing\neq K\subseteq X$ is convex and closed, and $K\cap\intefd{\Psi}\neq\varnothing$, then $K$ is left $D_\Psi$-Chebysh\"{e}v, and, for any $w\in K$ and any $x\in\intefd{\Psi}$, $w$ is the unique solution of
\begin{equation}
D_\Psi(y,z)+D_\Psi(z,x)\leq D_\Psi(y,x)\;\;\forall y\in K
\label{eqn.gen.pyth.ineq}
\end{equation}
(with respect to $z$) if{}f $w=\LPPP^{D_\Psi}_K(x)$. Furthermore, in `then' case of this `if{}f', if $K$ is affine, then $\leq$ in \eqref{eqn.gen.pyth.ineq} turns into $=$.
\end{proposition}

\begin{proposition}\label{prop.D.psi.projections.right}
\textup{\cite[Prop. 4.11]{MartinMarquez:Reich:Sabach:2012}}
If $(X,\n{\cdot}_X)$ is reflexive, $\Psi\in\pclg(X,\n{\cdot}_X)$ and $\efd(\Psi)=X$, $\Psi^\lfdual\in\pclg(X^\star,\n{\cdot}_{X^\star})$ is totally convex, $\varnothing\neq K\subseteq X$, and $\DG\Psi(K)$ is convex and closed, then $K$ is right $D_\Psi$-Chebysh\"{e}v, and, for any $w\in K$ and $x\in X$, $w$ is the unique solution of
\begin{equation}
D_\Psi(x,z)+D_\Psi(z,y)\leq D_\Psi(x,y)\;\;\forall y\in K
\label{eqn.gen.pyth.ineq.right}
\end{equation}
(with respect to $z$) if{}f $w=\RPPP^{D_\Psi}_K(x)$. Furthermore, in `then' case of this `if{}f', if $\DG\Psi(K)$ is affine, then $\leq$ in \eqref{eqn.gen.pyth.ineq.right} turns into $=$.
\end{proposition}

\begin{proposition}\label{prop.BBC.lemma}
\textup{\cite[Lem. 6.2]{Bauschke:Borwein:Combettes:2001}} 
Let $\Psi=\Psi_{1,\beta}:=\beta\n{\cdot}_X^{1/\beta}$, $\beta\in\,]0,1[$, for a reflexive $(X,\n{\cdot}_X)$. Then $\Psi_{1,\beta}$ is Euler--Legendre if{}f $(X,\n{\cdot}_X)$ is Gateaux differentiable and strictly convex. Furthermore, in such case $\Psi_{1,\beta}$ is also strictly convex on $\intefd{\Psi_{1,\beta}}=X$.
\end{proposition}

\begin{proposition}\label{prop.Alber}
\textup{\cite[\S7]{Alber:1993}(+\cite[I.3.1.g]{Cioranescu:1974})} If $(X,\n{\cdot}_X)$ is Gateaux differentiable, and $\Psi=\Psi_{1,\beta}:=\beta\n{\cdot}_X^{1/\beta}$, then $\DG\Psi_{1,\beta}(x)=\n{x}_X^{1/\beta-2}j(x)$, and
\begin{equation}
D_{\Psi_{1,\beta}}(x,y)=\beta\n{x}_X^{1/\beta}+(1-\beta)\n{y}_X^{1/\beta}-\n{y}_X^{1/\beta-2}\duality{x,j(y)}_{X\times X^\star}\in\RR^+
\label{eqn.Psi.1.beta}
\end{equation}
$\forall x,y\in X$, where $j(y)$ is defined as \textup{\cite[p. 35]{Klee:1953}} \textup{\cite[p. 211]{Vainberg:1961}} $z\in X^\star$ such that $\duality{y,z}_{X\times X^\star}=\n{y}_X\n{z}_{X^\star}$ and $\n{z}_{X^\star}=\n{y}_X$.
\end{proposition}

\begin{proposition}\label{prop.Resmerita.cor}
\textup{\cite[Cor. 4.4.(ii)]{Resmerita:2004}} 
If $(X,\n{\cdot}_X)$ is reflexive, strictly convex, Fr\'{e}chet differentiable, and has Radon--Riesz--Shmul'yan property,  $\varnothing\neq K\subseteq X$ is convex and closed, and $\Psi=\Psi_{\beta,\beta}:=\n{\cdot}_X^{1/\beta}$, $\beta\in\,]0,1[$, then $\LPPP^{D_{\Psi_{\beta,\beta}}}_K$ is norm-to-norm continuous on $\intefd{\Psi_{\beta,\beta}}=X$. 
\end{proposition}


\begin{proposition}\label{prop.Resmerita.total.convexity}
\textup{\cite[Thms. 3.1, 3.3]{Resmerita:2004}} 
If $(X,\n{\cdot}_X)$ is reflexive and $\beta\in\,]0,1[$, then $\Psi_{\beta,\beta}$ is totally convex if{}f $\Psi_{1,\beta}$ is totally convex if{}f $(X,\n{\cdot}_X)$ is strictly convex and has Radon--Riesz--Shmul'yan property. 
\end{proposition}

\begin{remark}\label{remark.complex}
If $(X,\n{\cdot}_X)$ is a Banach space over $\CC$, then  Propositions \ref{prop.D.psi.information}, \ref{prop.D.psi.projections}, \ref{prop.D.psi.projections.right}, \ref{prop.BBC.lemma}, \ref{prop.Alber}, \ref{prop.Resmerita.cor} hold under replacing $\duality{\cdot,\cdot}_{X\times X^\star}$ in Definition \ref{def.Bregman.reflexive} by $\re\duality{\cdot,\cdot}_{X\times X^\star}$. In Section \ref{section.bregman.nonreflexive} we will keep working with $(X,\n{\cdot}_X)$ over $\RR$, while in Section \ref{section.nonreflexive.applications} we will make use also of the case of $(X,\n{\cdot}_X)$ over $\CC$.
\end{remark}

\begin{remark}\label{remark.BCC.Resmerita.extension}
The proofs of Propositions \ref{prop.BBC.lemma}, \ref{prop.Resmerita.cor}, and \ref{prop.Resmerita.total.convexity} hold, without any additional alteration, under replacing $\Psi$ in each of these propositions by $\Psi_{\alpha,\beta}:=\frac{\beta}{\alpha}\n{\cdot}_X^{1/\beta}$, with $\beta\in\,]0,1[$ and $\alpha\in\,]0,\infty[$. ($\Psi_{{\alpha,\beta}}$ has appeared earlier in \cite[p. 616]{Iusem:GarcigaOtero:2001}.) In such case $\DG\Psi_{{\alpha,\beta}}(x)=\frac{1}{\alpha}\n{x}_X^{1/\beta-2}j(x)$, and $\forall x,y\in X$
\begin{equation}
D_{\Psi_{{\alpha,\beta}}}(x,y)=\frac{1}{\alpha}\left(\beta\n{x}_X^{1/\beta}+(1-\beta)\n{y}_X^{1/\beta}-\n{y}_X^{1/\beta-2}\duality{x,j(y)}_{X\times X^\star}\right).
\label{eqn.psi.varphi.alpha.beta}
\end{equation}
\end{remark}

\subsection{Extension to nonreflexive Banach spaces}\label{section.bregman.nonreflexive}

\begin{definition}\label{def.ell.Psi.information}
Let $(Y,\n{\cdot}_Y)$ be a Banach space, let $(X,\n{\cdot}_X)$ be a reflexive Banach space,  let $\varnothing\neq Z\subseteq Y$, and let $\ell:Z\ra\ell(Z)\subseteq X$. Then:
\begin{enumerate}[nosep,label=(\roman*)]
\item\label{def.ell.Psi.information.i} if $\varnothing\neq C\subseteq Z$ and $\ell(C)$ is convex (resp., closed; affine), then $C$ will be called \df{$\ell$-convex} (resp., \df{$\ell$-closed}; \df{$\ell$-affine}).
\end{enumerate}
If, furthermore, $\Psi\in\pclg(X,\n{\cdot}_X)$ is strictly convex on $\intefd{\Psi}$, and $\ell$ is a bijection such that $\ell(Z)\cap\intefd{\Psi}\neq\varnothing$, then: 
\begin{enumerate}[nosep,label=(\roman*)]
\setcounter{enumi}{1}
\item\label{def.ell.Psi.information.ii} an \df{extended Va\u{\i}nberg--Br\`{e}gman information} on Z is defined by
\begin{equation}
\hspace{-0.9cm}D_{\ell,\Psi}(\phi,\psi):=
        \left\{
                \begin{array}{ll}
                       \!\!D_\Psi(\ell(\phi),\ell(\psi))\!\!\! &\st(\phi,\psi)\in Z\times\ell^\inver(\ell(Z)\cap\intefd{\Psi})\\
                        \!\!\infty
												&\st\mbox{otherwise};
                \end{array}
        \right.\!\!\!
\label{eqn.generalised.Bregman}
\end{equation}
\item\label{def.ell.Psi.information.iii} a triple $(Z,\ell,\Psi)$ will be called an \df{extended Va\u{\i}nberg--Br\`{e}gman geometry}.
\end{enumerate}
\end{definition}

\begin{proposition}\label{prop.D.ell.psi.properties}
Let $(Y,\n{\cdot}_Y)$ be a Banach space, let $(X,\n{\cdot}_X)$ be a reflexive Banach space, $\varnothing\neq Z\subseteq Y$, $\ell:Z\ra X$, let $(Z,\ell,\Psi)$ be an extended Va\u{\i}nberg--Br\`{e}gman geometry. Then
\begin{enumerate}[nosep,label=(\roman*)]
\item\label{prop.D.ell.psi.properties.i} $D_{\ell,\Psi}$ is an information on $Z$ (justifying the use of a term `information' in Definition \ref{def.ell.Psi.information}.\ref{def.ell.Psi.information.ii}).
\end{enumerate}
If $\Psi$ is Euler--Legendre, $\psi\in\ell^\inver(\ell(Z)\cap\intefd{\Psi})$, and $\varnothing\neq C\subseteq Z$ is $\ell$-convex and $\ell$-closed, then:
\begin{enumerate}[nosep,label=(\roman*)]
\setcounter{enumi}{1}
\item\label{prop.D.ell.psi.properties.ii} the set $\arginff{\phi\in C}{D_{\ell,\Psi}(\phi,\psi)}$ is a singleton (with its element, denoted by $\LPPP^{D_{\ell,\Psi}}_C(\psi)$, equal to $\ell^\inver\circ\LPPP^{D_\Psi}_{\ell(C)}\circ\ell(\psi)$), i.e. $C$ is left $D_{\ell,\Psi}$-Chebysh\"{e}v;
\item\label{prop.D.ell.psi.properties.iii} $\omega=\LPPP^{D_{\ell,\Psi}}_C(\psi)$ if{}f $\omega$ is the unique solution of
\begin{equation}
D_{\ell,\Psi}(\phi,\zeta)+D_{\ell,\Psi}(\zeta,\psi)\leq D_{\ell,\Psi}(\phi,\psi)\;\;\forall\phi
\in C;
\label{eqn.gen.pyth.ineq.ell.Psi}
\end{equation}
\item\label{prop.D.ell.psi.properties.iv} if $C$ is $\ell$-affine, then $\leq$ in \eqref{eqn.gen.pyth.ineq.ell.Psi}, in `then' case of (iii), turns into $=$;
\item\label{prop.D.ell.psi.properties.v} if $\varnothing\neq K\subseteq X$ is convex and closed, and $\LPPP^{D_{\Psi}}_K$, $\ell$, and $\ell^\inver$ are norm-to-norm continuous, then $\LPPP^{D_{\ell,\Psi}}_C$ is norm-to-norm continuous on $Z$.
\end{enumerate}
If $\Psi^\lfdual\in\pclg(X^\star,\n{\cdot}_{X^\star})$ is totally convex, $\efd(\Psi)=X$, $\phi\in Z$, $\varnothing\neq C\subseteq Z$, and $C$ is $(\DG\Psi\circ\ell)$-convex and $(\DG\Psi\circ\ell)$-closed, then:
\begin{enumerate}[nosep,label=(\roman*)]
\setcounter{enumi}{5}
\item\label{prop.D.ell.psi.properties.vi} the set $\arginff{\phi\in C}{D_{\ell,\Psi}(\psi,\phi)}$ is a singleton (with its element, denoted by $\RPPP^{D_{\ell,\Psi}}_C(\psi)$, equal to $\ell^\inver\circ\RPPP^{D_\Psi}_{\ell(C)}\circ\ell(\psi)$), i.e. $C$ is right $D_{\ell,\Psi}$-Chebysh\"{e}v;
\item\label{prop.D.ell.psi.properties.vii} $\omega=\RPPP^{D_{\ell,\Psi}}_C(\phi)$ if{}f $\omega$ is the unique solution of
\begin{equation}
D_{\ell,\Psi}(\phi,\zeta)+D_{\ell,\Psi}(\zeta,\psi)\leq D_{\ell,\Psi}(\phi,\psi)\;\;\forall\psi
\in C;
\label{eqn.gen.pyth.ineq.ell.Psi.right}
\end{equation}
\item\label{prop.D.ell.psi.properties.viii} if $C$ is $(\DG\Psi\circ\ell)$-affine, then $\leq$ in \eqref{eqn.gen.pyth.ineq.ell.Psi.right}, in `then' case of (vii), turns into $=$.
\end{enumerate}
\end{proposition}
\begin{proof}
\ref{prop.D.ell.psi.properties.i} and \ref{prop.D.ell.psi.properties.ii}--\ref{prop.D.ell.psi.properties.iv} (resp., \ref{prop.D.ell.psi.properties.vi}--\ref{prop.D.ell.psi.properties.viii}) follow from Propositions \ref{prop.D.psi.information} and \ref{prop.D.psi.projections} (resp., \ref{prop.D.psi.projections.right}), combined with bijectivity of $\ell$, while \ref{prop.D.ell.psi.properties.v} follows from bijectivity of $\ell$ and compositionality of norm-to-norm continuous maps.
\end{proof}

\begin{proposition}\label{prop.left.pythagorean.info.Psi.alpha.beta}
If $(X,\n{\cdot}_X)$ is a strictly convex, Gateaux differentiable, reflexive Banach space, $(Y,\n{\cdot}_Y)$ is a Banach space, $\varnothing\neq Z\subseteq Y$, $\Psi=\Psi_{\alpha,\beta}:=\frac{\beta}{\alpha}\n{\cdot}_X^{1/\beta}$, $\beta\in\,]0,1[$, $\alpha\in\,]0,\infty[$, $\ell:Z\ra\ell(Z)\subseteq X$ is a bijection, $\varnothing\neq C\subseteq Z$ is $\ell$-convex and $\ell$-closed, then:
\begin{enumerate}[nosep,label=(\roman*)]
\item\label{prop.left.pythagorean.info.Psi.alpha.beta.i} $\intefd{\Psi_{\alpha,\beta}}=X$;
\item\label{prop.left.pythagorean.info.Psi.alpha.beta.ii} $D_{\ell,\Psi_{\alpha,\beta}}$ is an information on $Z$;
\item\label{prop.left.pythagorean.info.Psi.alpha.beta.iii} $C$ is left $D_{\ell,\Psi_{\alpha,\beta}}$-Chebysh\"{e}v;
\item\label{prop.left.pythagorean.info.Psi.alpha.beta.iv} $\omega=\LPPP^{D_{\ell,\Psi_{\alpha,\beta}}}_C(\psi)$ if{}f $\omega$ is the unique solution of
\begin{equation}
D_{\ell,\Psi_{\alpha,\beta}}(\phi,\zeta)+D_{\ell,\Psi_{\alpha,\beta}}(\zeta,\psi)\leq D_{\ell,\Psi_{\alpha,\beta}}(\phi,\psi)\;\;\forall (\phi,\psi)\in C\times Z;
\label{eqn.gen.pyth.ineq.Psi.alpha.beta}
\end{equation}
\item\label{prop.left.pythagorean.info.Psi.alpha.beta.v} if $C$ is $\ell$-affine, then $\leq$ in \eqref{eqn.gen.pyth.ineq.Psi.alpha.beta}, in `then' case of \ref{prop.left.pythagorean.info.Psi.alpha.beta.iv}, turns into $=$.
\end{enumerate}
If, furthermore, ($X,\n{\cdot}_X)$ is Fr\'{e}chet differentiable, $\varnothing\neq\widetilde{C}\subseteq Z$, and $\widetilde{C}$ is $(\DG\Psi_{\alpha,\beta}\circ\ell)$-convex and $\ell$-closed, then:
\begin{enumerate}[nosep,label=(\roman*)]
\setcounter{enumi}{5}
\item\label{prop.left.pythagorean.info.Psi.alpha.beta.vi} if $(X,\n{\cdot}_X)$  has Radon--Riesz--Shm\-ul'yan property and $\ell$ is a norm-to-norm homeomorphism, then $\LPPP^{D_{\ell,\Psi_{\alpha,\beta}}}_C$ is norm-to-norm continuous on $Z$;
\item\label{prop.left.pythagorean.info.Psi.alpha.beta.vii}$\widetilde{C}$ is right $D_{\ell,\Psi_{\alpha,\beta}}$-Chebysh\"{e}v;
\item\label{prop.left.pythagorean.info.Psi.alpha.beta.viii} $\omega=\RPPP^{D_{\ell,\Psi_{\alpha,\beta}}}_{\widetilde{C}}(\phi)$ if{}f $\omega$ is the unique solution of
\begin{equation}
D_{\ell,\Psi_{\alpha,\beta}}(\phi,\zeta)+D_{\ell,\Psi_{\alpha,\beta}}(\zeta,\psi)\leq D_{\ell,\Psi_{\alpha,\beta}}(\phi,\psi)\;\;\forall(\phi,\psi)\in Z\times\widetilde{C};
\label{eqn.gen.pyth.ineq.Psi.alpha.beta.right}
\end{equation}
\item\label{prop.left.pythagorean.info.Psi.alpha.beta.ix} if $\widetilde{C}$ is $(\DG\Psi_{\alpha,\beta}\circ\ell)$-affine, then $\leq$ in \eqref{eqn.gen.pyth.ineq.Psi.alpha.beta.right}, in `then' case of \ref{prop.left.pythagorean.info.Psi.alpha.beta.viii}, turns into $=$.
\end{enumerate}
\end{proposition}
\begin{proof}
\ref{prop.left.pythagorean.info.Psi.alpha.beta.i} follows from the finiteness of the values of $\Psi_{\alpha,\beta}$; \ref{prop.left.pythagorean.info.Psi.alpha.beta.ii}--\ref{prop.left.pythagorean.info.Psi.alpha.beta.ix} follow from Propositions \ref{prop.BBC.lemma}, \ref{prop.Resmerita.cor}, and \ref{prop.Resmerita.total.convexity}, combined with Remark \ref{remark.BCC.Resmerita.extension} and Proposition \ref{prop.D.ell.psi.properties}. For \ref{prop.left.pythagorean.info.Psi.alpha.beta.vi}--\ref{prop.left.pythagorean.info.Psi.alpha.beta.ix} it is necessary to use \cite[Thm. 3.8]{Anderson:1960}: if $(X,\n{\cdot}_X)$ is reflexive and Fr\'{e}chet differentiable then $(X^\star,\n{\cdot}_{X^\star})$ is strictly convex and has Radon--Riesz--Shmul'yan property. By \cite[Cor. 4.12]{Cudia:1964}, Fr\'{e}chet differentiability of $(X,\n{\cdot}_X)$ is equivalent with norm-to-norm continuity of $j$, and this implies norm-to-norm continuity of $\DG\Psi_{\alpha,\beta}$. Hence, $(\DG\Psi_{\alpha,\beta}\circ\ell)$-closed sets coincide with $\ell$-closed sets.
\end{proof}

\begin{remark}
If $(X,\n{\cdot}_X)$ is a Hilbert space, then Proposition \ref{prop.left.pythagorean.info.Psi.alpha.beta} features metric projections on $(X,\n{\cdot}_X)$, arising as a special case of $D_{\Psi_{\alpha,\beta}}$-projections for $(\alpha,\beta)=(1,\frac{1}{2})$.
\end{remark}

\begin{remark}\label{remark.Psi.varphi}
The role of $\Psi=\Psi_{\alpha,\beta}$ in this work is to provide explicit quantitative models of general results based on Proposition \ref{prop.D.ell.psi.properties}. By this reason, we distinguish between the cases \ref{prop.Wstar.gamma.alpha.beta.i}--\ref{prop.Wstar.gamma.alpha.beta.iii} and \ref{prop.Wstar.gamma.alpha.beta.iv} (resp., \ref{prop.JBW.D.alpha.beta.gamma.i}--\ref{prop.JBW.D.alpha.beta.gamma.iii} and \ref{prop.JBW.D.alpha.beta.gamma.iv}--\ref{prop.JBW.D.alpha.beta.gamma.vi}; \ref{prop.noncommutative.Orlicz.D.ell.Psi.i}--\ref{prop.noncommutative.Orlicz.D.ell.Psi.iii} and \ref{prop.noncommutative.Orlicz.D.ell.Psi.iv}--\ref{prop.noncommutative.Orlicz.D.ell.Psi.v}; \ref{prop.lozanovskii.D.ell.Psi.ii}--\ref{prop.lozanovskii.D.ell.Psi.iv} and \ref{prop.lozanovskii.D.ell.Psi.v}) in Proposition \ref{prop.Wstar.gamma.alpha.beta} (resp., \ref{prop.JBW.D.alpha.beta.gamma}; \ref{prop.noncommutative.Orlicz.D.ell.Psi}; \ref{prop.lozanovskii.D.ell.Psi}). With an exception of explicit formulas for $D_{\ell,\Psi_{\alpha,\beta}}$, the results in Propositions \ref{prop.Wstar.gamma.alpha.beta}.\ref{prop.Wstar.gamma.alpha.beta.iv}, \ref{prop.JBW.D.alpha.beta.gamma}.\ref{prop.JBW.D.alpha.beta.gamma.iv}--\ref{prop.JBW.D.alpha.beta.gamma.vi}, \ref{prop.noncommutative.Orlicz.D.ell.Psi}.\ref{prop.noncommutative.Orlicz.D.ell.Psi.iv}--\ref{prop.noncommutative.Orlicz.D.ell.Psi.v}, \ref{cor.commutative.Orlicz}.\ref{cor.commutative.Orlicz.i}, \ref{prop.gen.spin.factor.D.alpha.beta.gamma}, \ref{prop.lozanovskii.D.ell.Psi}.\ref{prop.lozanovskii.D.ell.Psi.v}, and \ref{prop.nc.lozanovskii.D.ell.Psi}.\ref{prop.nc.lozanovskii.D.ell.Psi.ii} do not depend  on the particular form of $\Psi=\Psi_{\alpha,\beta}$, but only on the fact that its properties (including the properties of $\LPPP^{D_\Psi}$ and $\RPPP^{D_\Psi}$) are determined, via Propositions \ref{prop.BBC.lemma}--\ref{prop.Resmerita.total.convexity}, by geometric properties of the norm of an underlying reflexive Banach space. Hence, it is natural to ask about more general class of functions on reflexive Banach spaces $(X,\n{\cdot}_X)$, which would allow for a suitable control of the properties of $D_\Psi$, $\LPPP^{D_\Psi}$, and $\RPPP^{D_\Psi}$, by means of the differentiability and convexity properties of $(X,\n{\cdot}_X)$. This can be achieved by consideration of a class of functions \cite[p. 200]{Asplund:1967} $\Psi_\varphi(x):=\int_0^{\n{x}_X}\dd t\,\varphi(t)$, where $\varphi:\RR^+\ra\RR^+$ is strictly increasing, continuous, $\varphi(0)=0$, and $\lim_{t\ra\infty}\varphi(t)=\infty$ \cite[p. 407]{Beurling:Livingston:1962}. (In particular, $\Psi_{\alpha,\beta}=\Psi_{\varphi_{\alpha,\beta}}$ with $\varphi_{\alpha,\beta}(t)=\frac{1}{\alpha}t^{1/\beta-1}$.) However, since it requires us to develop suitable generalisations of the convex analytic results contained in Propositions \ref{prop.BBC.lemma}--\ref{prop.Resmerita.cor},\footnote{\cite[Thm. 3.1]{Resmerita:2004} already provides a corresponding generalisation of Proposition 2.\ref{prop.Resmerita.total.convexity}.} this will be provided in another paper  \cite{Kostecki:2025}.
\end{remark}

\section{Application to some base normed spaces}\label{section.nonreflexive.applications}

For a detailed account of the integration theory on W$^*$-algebras (resp., JBW-algebras; radially compact base normed spaces in spectral duality), we refer to \cite{Stratila:1981,Falcone:Takesaki:2001,Takesaki:2003,Kostecki:2013,Dodds:dePagter:Sukochev:2023} (resp., \cite{Iochum:1984,Abdullaev:1984,Ayupov:1986,Iochum:1986}; \cite{Alfsen:Shultz:1976,Alfsen:Shultz:1979,Yadgorov:1989,Tikhonov:1993,Berdikulov:2005,Berdikulov:2006}).

\subsection{Noncommutative $L_p$ spaces and Mazur maps}

\begin{definition}\label{def.noncommutative.mazur} 
\textup{\cite[p. 58]{Raynaud:2002}} For any W$^*$-algebra $\N$, and $\gamma_1,\gamma_2\in\,]0,\infty[$, a \df{noncommutative Mazur map} is defined as 
\begin{equation}
\ell_{\gamma_1,\gamma_2}:L_{1/\gamma_1}(\N)\ni x=u_x\ab{x}\mapsto u_x\ab{x}^{\gamma_2/\gamma_1}\in L_{1/\gamma_2}(\N),
\label{eqn.noncommutative.mazur}
\end{equation}
where $x=u_x\ab{x}$ is the unique polar decomposition of $x$, while the meaning of the symbol `$\ab{x}^{\gamma_2/\gamma_1}$' is given in \textup{\cite[p. 196]{Falcone:Takesaki:2001}}. Also, $\ell_{\gamma_2}:=\ell_{1,\gamma_2}$.
\end{definition}

\begin{proposition}\label{prop.Wstar.gamma.alpha.beta} 
Let $\N$ be a W$^*$-algebra, $\gamma,\beta\in\,]0,1[$, $\lambda,\alpha\in\,]0,\infty[$, $\varnothing\neq C\subseteq\N_{\star}$, and let $\Psi\in\pclg(L_{1/\gamma}(\N),\n{\cdot}_{1/\gamma})$ be strictly convex on $\efd(\Psi)=L_{1/\gamma}(\N)$. Then:
\begin{enumerate}[nosep,label=(\roman*)]
\item\label{prop.Wstar.gamma.alpha.beta.i}  $D_{\lambda\ell_\gamma,\Psi}$ is an information on $\N_\star$;
\item\label{prop.Wstar.gamma.alpha.beta.ii}  if $\Psi$ is Euler--Legendre and $C$ is $\lambda\ell_\gamma$-convex and closed, then $C$ is left $D_{\lambda\ell_\gamma,\Psi}$-Chebysh\"{e}v, and $\omega=\LPPP^{D_{\lambda\ell_\gamma,\Psi}}_C(\psi)$ if{}f $\omega$ is a unique solution of
\begin{equation}
D_{\lambda\ell_\gamma,\Psi}(\phi,\zeta)+D_{\lambda\ell_\gamma,\Psi}(\zeta,\psi)\leq D_{\lambda\ell_\gamma,\Psi}(\phi,\psi)\;\forall(\phi,\psi)\in C\times\N_\star,
\label{eqn.gen.pyth.ineq.Psi.alpha.beta.gamma}
\end{equation}
with $\leq$ replaced by $=$, in `then' case of this `if{}f', if $C$ is $\lambda\ell_\gamma$-affine;
\item\label{prop.Wstar.gamma.alpha.beta.iii}  if $\Psi^\lfdual\in\pclg(L_{1/(1-\gamma)}(\N),\n{\cdot}_{1/(1-\gamma)})$ is totally convex, $C$ is $(\DG\Psi\circ\lambda\ell_\gamma)$-convex and $(\DG\Psi\circ\lambda\ell_\gamma)$-closed, then $C$ is right $D_{\lambda\ell_\gamma,\Psi}$-Chebysh\"{e}v, and $\omega=\RPPP^{D_{\lambda\ell_\gamma,\Psi}}_C(\phi)$ if{}f $\omega$ is a unique solution of
\begin{equation}
D_{\lambda\ell_\gamma,\Psi}(\phi,\zeta)+D_{\lambda\ell_\gamma,\Psi}(\zeta,\psi)\leq D_{\lambda\ell_\gamma,\Psi}(\phi,\psi)\;\forall(\phi,\psi)\in\N_\star\times C,
\label{eqn.gen.pyth.ineq.Psi.alpha.beta.gamma.right}
\end{equation}
with $\leq$ replaced by $=$, in `then' case of this `if{}f', if $C$ is $(\DG\Psi\circ\lambda\ell_\gamma)$-affine;
\item\label{prop.Wstar.gamma.alpha.beta.iv} if $\Psi=\Psi_{\alpha,\beta}=\frac{\beta}{\alpha}\n{\cdot}_{1/\gamma}^{1/\beta}$, then:
\begin{enumerate}[nosep,label=\alph*)]
\item\label{prop.Wstar.gamma.alpha.beta.iv.a}the assumptions, and thus the conclusions, of \ref{prop.Wstar.gamma.alpha.beta.i}--\ref{prop.Wstar.gamma.alpha.beta.iii} hold for $D_{\lambda\ell_\gamma,\Psi}=D_{\lambda\ell_\gamma,\Psi_{{\alpha,\beta}}}$, and the $(\DG\Psi_{\alpha,\beta}\circ\lambda\ell_\gamma)$-closed sets are closed; 
\item\label{prop.Wstar.gamma.alpha.beta.iv.b} $\LPPP^{D_{\lambda\ell_\gamma,\Psi_{\alpha,\beta}}}_C$ is norm-to-norm continuous on $(\N_\star,\n{\cdot}_{\N_\star})$;
\item\label{prop.Wstar.gamma.alpha.beta.iv.c} $\forall(\phi,\psi)\in\N_\star\times\N_\star$ 
\begin{align}
D_{\lambda\ell_\gamma,\Psi_{{\alpha,\beta}}}(\phi,\psi)=&\,\textstyle\frac{\lambda^{1/\beta}}{\alpha}\left(\beta\n{\phi}^{\gamma/\beta}_1+(1-\beta)\n{\psi}_1^{\gamma/\beta}
-\n{\psi}_1^{\frac{\gamma}{\beta}-1}\,\re\!\!\int u_\phi\ab{\phi}^\gamma u_\psi\ab{\psi}^{1-\gamma}\right)\in\RR^+,
\label{eqn.lambda.gamma.alpha.beta}
\end{align}
where the symbol `$\int$' is understood in the sense of \textup{\cite[Eqn. (3.12')]{Falcone:Takesaki:2001}}.
\end{enumerate}
\end{enumerate}
\end{proposition}
\begin{proof}\ \\
\begin{enumerate}[nosep]
\item[(i)--(iii)]  By \cite[Lem. 3.2]{Raynaud:2002}, $\ell_\gamma$ is a norm-to-norm homeomorphism from $(\N_\star,\n{\cdot}_1)\iso(L_1(\N),\n{\cdot}_1)$ to $(L_{1/\gamma}(\N),\n{\cdot}_{1/\gamma})$ for any $\gamma\in\,]0,1[$. The rest follows from Proposition \ref{prop.D.ell.psi.properties}.
\item[(iv)] Since $\intefd{\Psi_{\alpha,\beta}}=L_{1/\gamma}(\N)$, we have $(\lambda\ell_\gamma)^\inver(\intefd{\Psi_{\alpha,\beta}})=\N_\star$. \eqref{eqn.lambda.gamma.alpha.beta} follows from \eqref{eqn.psi.varphi.alpha.beta} by a direct calculation, using \cite[Lem. 3.1]{Kosaki:1984:uniform}
\begin{equation}
L_{1/\gamma}(\N)\ni x\mapsto j(x)=\n{x}^{2-1/\gamma}_{1/\gamma}u_x\ab{x}^{1/\gamma-1}\in L_{1/(1-\gamma)}(\N)
\end{equation}
(the latter following from \cite[Prop. 24]{Terp:1981} \cite[p. 162]{Hilsum:1981}; cf. also \cite[Eqn. (11)]{Jencova:2005}), with $x=u_x\ab{x}$. For any $\gamma\in\,]0,1[$, $(L_{1/\gamma}(\N),\n{\cdot}_{1/\gamma})$ is uniformly convex \cite[Lems. 8.1, 8.2]{Masuda:1983} \cite[Thm. 5.3]{Fack:Kosaki:1986}. Due to the Banach duality \cite[Thm. 3.4.3]{Kosaki:1980:PhD} \cite[Thm. 10.(2)]{Hilsum:1981} \cite[Thm. 32.(2)]{Terp:1981}
\begin{equation}
(L_{1/\gamma}(\N),\n{\cdot}_{1/\gamma})^\star\iso(L_{1/(1-\gamma)}(\N),\n{\cdot}_{1/(1-\gamma)})\;\forall\gamma\in\,]0,1[,
\end{equation}
this is equivalent, by \cite[Cor. (p. 647)]{Shmulyan:1940:diff} \cite[Thm. 6.7]{Day:1944}, with uniform Fr\'{e}chet differentiability of $(L_{1/\gamma}(\N),\n{\cdot}_{1/\gamma})$ for $\gamma\in\,]0,1[$. Uniform convexity of a Banach space entails its Radon--Riesz--Shmul'yan property, strict convexity, and reflexivity \cite[Thm. 2]{Milman:1938}, while uniform Fr\'{e}chet differentiability entails Fr\'{e}chet differentiability and Gateaux differentiability. Hence, for any $\gamma\in\,]0,1[$, $\Psi_{\alpha,\beta}$ is Euler--Legendre on $(L_{1/\gamma}(\N),\n{\cdot}_{1/\gamma})$, and ${\Psi_{\alpha,\beta}}^\lfdual$ is totally convex on $(L_{1/\gamma}(\N),\n{\cdot}_{1/\gamma})$, by means of Propositions \ref{prop.BBC.lemma}, \ref{prop.Resmerita.total.convexity}, and Remark \ref{remark.BCC.Resmerita.extension}. The rest follows from Proposition \ref{prop.left.pythagorean.info.Psi.alpha.beta}.\ref{prop.left.pythagorean.info.Psi.alpha.beta.ii}--\ref{prop.left.pythagorean.info.Psi.alpha.beta.ix} and Remark \ref{remark.complex}.
\end{enumerate}
\end{proof}

\begin{corollary}\label{cor.d.gamma}
\begin{enumerate}[nosep,label=(\roman*)]
\item\label{cor.d.gamma.i} $D_{\lambda\ell_\gamma,\Psi_{{\alpha,\beta}}}=D_{\ell_\gamma,\Psi_{{\alpha\lambda^{-1/\beta},\beta}}}$.
\item\label{cor.d.gamma.ii} For $\lambda=1$, $\beta=\gamma$, $\alpha=\gamma(1-\gamma)$, we obtain $\Psi_{{\gamma(1-\gamma),\gamma}}(x)=\frac{1}{1-\gamma}\n{x}_{1/\gamma}^{1/\gamma}$ $\forall x\in L_{1/\gamma}(\N)$, and $\forall\phi,\psi\in\N_\star$
\begin{align}
\hspace{-1.2cm}D_{\ell_\gamma,\Psi_{{\gamma(1-\gamma),\gamma}}}(\phi,\psi)&=D_{\frac{1}{\gamma}\ell_\gamma,\Psi_{{\gamma^{1-1/\gamma}(1-\gamma),\gamma}}}(\phi,\psi)
\label{eqn.psi.gamma.equals.psi.gamma}
\\
&=\frac{\n{\phi}_1}{1-\gamma}+\frac{\n{\psi}_1}{\gamma}+\frac{\re\int u_\phi\ab{\phi}^\gamma u_\psi\ab{\psi}^{1-\gamma}}{\gamma(1-\gamma)}=:D_\gamma(\phi,\psi).
\label{eqn.d.gamma}
\end{align}
\item\label{cor.d.gamma.iii} If $\N$ is semifinite and $\tau$ is a normal, faithful, and semifinite trace on $\N$ (e.g., if $\N=\BH$ and $\tau=\tr_\H$), then $\forall(\phi,\psi)\in{\N_\star}^+$
\begin{equation}
D_{\lambda\ell_\gamma,\Psi_{{\alpha,\beta}}}(\phi,\psi)=\textstyle\frac{\lambda^{1/\beta}}{\alpha}(\beta(\tau(\rho_\phi))^{\gamma/\beta}+(1-\beta)(\tau(\rho_\psi))^{\gamma/\beta}-(\tau(\rho_\psi))^{\frac{\gamma}{\beta}-1}\tau(\rho_\phi^\gamma{\rho_\psi}^{1-\gamma})),
\end{equation}
where $\phi=:\tau(\rho_\phi\,\cdot\,)$ and $\psi=:\tau(\rho_\psi\,\cdot\,)$. In particular, if $\psi,\phi\in(S(\N_\star,\n{\cdot}_1))^+$, then
\begin{equation}
D_{\lambda\ell_\gamma,\Psi_{{\alpha,\beta}}}(\phi,\psi)=\textstyle\frac{\lambda^{1/\beta}}{\alpha}(1-\tau(\rho_\phi^\gamma{\rho_\psi}^{1-\gamma})).
\end{equation}
\end{enumerate}
\end{corollary}
\begin{proof}
Follows from \eqref{eqn.lambda.gamma.alpha.beta} by a direct calculation.
\end{proof}

\begin{remark}\label{remark.relation.with.jencova.results}
Identification of $D_\gamma$ as $D_{\ell_\gamma,\Psi_{{\gamma(1-\gamma),\gamma}}}$, provided in Corollary \ref{cor.d.gamma}.\ref{cor.d.gamma.ii}, is new. Up to reformulation in weight-independent terms, provided in \cite[Eqn. (41)]{Kostecki:2011:OSID}, the formula \eqref{eqn.d.gamma} was obtained in \cite[\S8]{Jencova:2005} (cf. also \cite[Eqn. (42)]{Ojima:2004}) as $D_\Psi(\frac{1}{\gamma}\ell_\gamma(\phi),\frac{1}{\gamma}\ell_\gamma(\psi))$ with $\Psi$ equal to $\Psi_{{\gamma^{1-1/\gamma}(1-\gamma),\gamma}}$ (however, it was not identified there as an example of $\Psi_{\alpha,\beta}$, although the corresponding $D_\Psi$ was explicitly identified as a Va\u{\i}nberg--Br\`{e}gman functional). (For $\N$ of type I$_n$ with $n\in\NN$, $D_\gamma$ was introduced in \cite[Eqn. (9)]{Hasegawa:1993}, while for commutative $\N$ it was introduced in \cite[Rem. 2.5]{Liese:Vajda:1987}.) Hence, Propositions \ref{prop.Wstar.gamma.alpha.beta}.\ref{prop.Wstar.gamma.alpha.beta.iii} and \ref{prop.Wstar.gamma.alpha.beta}.\ref{prop.Wstar.gamma.alpha.beta.iv}.\ref{prop.Wstar.gamma.alpha.beta.iv.a} provide a generalisation of \cite[Props. 8.1.(i)--(ii), 8.2.(ii)]{Jencova:2005} to pairs $(\ell_\gamma,\Psi_{\alpha,\beta})$ with any $(\alpha,\beta,\gamma)\in\,]0,\infty[\,\times\,]0,1[\,^2$, not necessarily $(\gamma(1-\gamma),\gamma,\gamma)$.
\end{remark}

\subsection{Nonassociative $L_p$ spaces and Mazur maps}

\begin{definition}\label{def.nonassociative.Mazur}
Let $A$ be a semifinite JBW-algebra with a Jordan product $\jordan$ and a unit $\II$, let $\tau$ be a faithful normal semifinite trace on $A$, $\gamma_1,\gamma_2\in\,]0,\infty[$. We define a \df{nonassociative Mazur map} as
\begin{equation}
\ell_{\gamma_1,\gamma_2}:L_{1/\gamma_1}(A,\tau)\ni x=s_x\jordan\ab{x}\mapsto s_x\jordan\ab{x}^{\gamma_2/\gamma_1}\in L_{1/\gamma_2}(A,\tau),
\end{equation}
where $x=s_x\jordan\ab{x}$ is a polar decomposition with $s_x\in A$ such that $s_x^2=\II$. Also, $\ell_{\gamma_2}:=\ell_{1,\gamma_2}$.
\end{definition}

\begin{proposition}\label{prop.na.l.gamma.hoelder}
Let $A$ be a semifinite JBW-algebra with a unit $\II$, let $\tau$ be a faithful normal semifinite trace on $A$, and let $\gamma,\gamma_1,\gamma_2\in\,]0,1]$. Then $\ell_{\gamma_1,\gamma_2}$ is $\min\{\frac{\gamma_2}{\gamma_1},1\}$-Lipschitz--H\"{o}lder continuous on $(B(L_{1/\gamma_1}(A,\tau),\n{\cdot}_{1/\gamma_1}))^+:=\{x\geq0\mid x\in B(L_{1/\gamma_1}(A,\tau),\n{\cdot}_{1/\gamma_1})\}$. In particular,
\begin{enumerate}[nosep,label=(\roman*)]
\item\label{prop.na.l.gamma.hoelder.i} $(\ell_\gamma)^\inver$ is Lipschitz continuous on $(B(L_{1/\gamma},\n{\cdot}_{1/\gamma}))^+$;
\item\label{prop.na.l.gamma.hoelder.ii} $\ell_\gamma$ is $\gamma$-Lipschitz--H\"{o}lder continuous on $(B(A_\star,\n{\cdot}_1))^+$.
\end{enumerate}
\end{proposition}
\begin{proof}
Any faithful normal semifinite trace $\bar{\tau}$ on a reversible JW-algebra $J$ can be extended to a faithful normal semifinite trace $\tilde{\tau}$ on an enveloping von Neumann algebra $\tilde{J}$ of $J$ \cite[Thm. 2]{Ayupov:1982:extension}. The type of $\tilde{J}$ is the same as the type of $J$ \cite[Thm. 8]{Ayupov:1982:extension}. Let now $J$ be a reversible JW-algebra that is a JBW-subalgebra of $A$, generated by $\II$ and $z,w\in A$. Given any formula of inequality involving $\bar{\tau}$ and $\{z,w,\II\}$, this formula holds if it is true under replacing $(J,\bar{\tau})$ by $(\tilde{J},\tilde{\tau})$ \cite[Rem. (p. 94)]{Ayupov:1986}. The inequality formula of $t$-Lipschitz--H\"{o}lder continuity of $\ell_{\gamma_1,\gamma_2}$ is: $\exists c>0$ $\forall x,y\in (B(L_{1/\gamma_1}(A,\tau),\n{\cdot}_{1/\gamma_1}))^+$ 
\begin{equation}
\left(\tau\left(\ab{x^{\gamma_2/\gamma_1}-y^{\gamma_2/\gamma_1}}^{1/\gamma_2}\right)\right)^{\gamma_2}\leq c(\tau(\ab{x-y}^{1/\gamma_1}))^{t\gamma_1}.
\end{equation}
Since $x,y\in A$, the result follows from the fact that the noncommutative Mazur map $\ell_{\gamma_1,\gamma_2}$ is $\min\{\frac{\gamma_2}{\gamma_1},1\}$-Lipschitz--H\"{o}lder continuous on $B(L_{1/\gamma_1}(\N),$ $\n{\cdot}_{1/\gamma_1})$ for any W$^*$-algebra $\N$ \cite[Thm. (p. 37)]{Ricard:2015}. (ii) also follows directly from \cite[Prop. 9.(ii)]{Iochum:1986}.
\end{proof}

\begin{proposition}\label{prop.JBW.D.alpha.beta.gamma}
Let $A$ be a semifinite JBW-algebra with  a Jordan product $\jordan$, let $\tau$ be a faithful normal semifinite trace on $A$,  $\varnothing\neq C\subseteq A_\star$, $\gamma,\beta\in\,]0,1[$, $\lambda,\alpha\in\,]0,\infty[$. Let $\Psi\in\pclg(L_{1/\gamma}(A,\tau),\n{\cdot}_{1/\gamma})$ be strictly convex on $\efd(\Psi)=L_{1/\gamma}(A,\tau)$. Then:
\begin{enumerate}[nosep,label=(\roman*)]
\item\label{prop.JBW.D.alpha.beta.gamma.i} $D_{\lambda\ell_\gamma,\Psi}$ is an information on $A_\star$;
\item\label{prop.JBW.D.alpha.beta.gamma.ii} if $\Psi$ is Euler--Legendre and $C$ is $\lambda\ell_\gamma$-convex and $\lambda\ell_\gamma$-closed, then $C$ is left $D_{\lambda\ell_\gamma,\Psi}$-Chebysh\"{e}v, and $\omega=\LPPP^{D_{\lambda\ell_\gamma,\Psi}}_C(\psi)$ if{}f $\omega$ is a unique solution of \eqref{eqn.gen.pyth.ineq.Psi.alpha.beta.gamma} under replacement of $\N_\star$ by $A_\star$, and, in `then' case of this `if{}f', with $\leq$ replaced by $=$ if $C$ is $\lambda\ell_\gamma$-affine;
\item\label{prop.JBW.D.alpha.beta.gamma.iii} if $\Psi^\lfdual\in\pclg(L_{1/(1-\gamma)}(\N),\n{\cdot}_{1/(1-\gamma)})$ is totally convex, and $C$ is $(\DG\Psi\circ\lambda\ell_\gamma)$-convex and $(\DG\Psi\circ\lambda\ell_\gamma)$-closed, then $C$ is right $D_{\lambda\ell_\gamma,\Psi}$-Chebysh\"{e}v, and $\omega=\RPPP^{D_{\lambda\ell_\gamma,\Psi}}_C(\phi)$ if{}f $\omega$ is a unique solution of \eqref{eqn.gen.pyth.ineq.Psi.alpha.beta.gamma.right} under replacement of $\N_\star$ by $A_\star$, and, in `then' case of this `if{}f', with $\leq$ replaced by $=$ if $C$ is $(\DG\Psi\circ\lambda\ell_\gamma)$-affine;
\item\label{prop.JBW.D.alpha.beta.gamma.iv} if $\Psi=\Psi_{\alpha,\beta}=\frac{\beta}{\alpha}\n{\cdot}_{1/\gamma}^{1/\beta}$, then $(\DG\Psi_{\alpha,\beta}\circ\lambda\ell_\gamma)$-closed sets coincide with $\lambda\ell_\gamma$-closed sets (as well as with closed sets, if they are the subsets of $(B(A_\star,\n{\cdot}_1))^+$), and the assumptions (and thus the conclusions) of \ref{prop.JBW.D.alpha.beta.gamma.i}--\ref{prop.JBW.D.alpha.beta.gamma.iii} hold for $D_{\lambda\ell_\gamma,\Psi_{{\alpha,\beta}}}=D_{\ell_\gamma,\Psi_{{\alpha\lambda^{-1/\beta},\beta}}}$, where 
\begin{align}
\hspace{-0.5cm}D_{\lambda\ell_\gamma,\Psi_{\alpha,\beta}}(\omega,\phi)=&\,\textstyle\frac{\lambda^{1/\beta}}{\alpha}(\beta(\tau(\rho_\omega))^{\gamma/\beta}+(1-\beta)(\tau(\rho_\phi))^{\gamma/\beta}\nonumber\\
&-(\tau(\rho_\phi))^{\gamma/\beta-1}\tau((s_\omega\jordan\ab{\rho_\omega}^\gamma)\jordan(s_\phi\jordan\ab{\rho_\phi}^{1-\gamma})))\in\RR^+
\end{align}
$\forall\omega,\phi\in A_\star$ such that $\omega=\tau(\rho_\omega\jordan\,\cdot\,)$ and $\phi=\tau(\rho_\phi\jordan\,\cdot\,)$;
\item\label{prop.JBW.D.alpha.beta.gamma.v} if $\varnothing\neq K\subseteq L_{1/\gamma}(A,\tau)$ is convex and closed, then $\LPPP^{D_{\Psi_{\alpha,\beta}}}_K$ is norm-to-norm continuous on $(L_{1/\gamma}(A,\tau),\n{\cdot}_{1/\gamma})$;
\item\label{prop.JBW.D.alpha.beta.gamma.vi} if $C\subseteq(B(A_\star,\n{\cdot}_1))^+$ is $\lambda\ell_\gamma$-convex and closed, then $\LPPP^{D_{\lambda\ell_\gamma,\Psi_{\alpha,\beta}}}_C$ is norm-to-norm continuous on $(B(A_\star,\n{\cdot}_1))^+$.
\end{enumerate}
\end{proposition}
\begin{proof}
Due to the Banach space duality \cite[Thm. 2.1.10]{Abdullaev:1984} \cite[Thm. V.3.2]{Iochum:1984}
\begin{equation}
(L_{1/\gamma}(A,\tau),\n{\cdot}_{1/\gamma})^\star\iso (L_{1/(1-\gamma)}(A,\tau),\n{\cdot}_{1/(1-\gamma)})\;\forall\gamma\in\,]0,1[,
\end{equation}
uniform convexity of $(L_{1/\gamma}(A,\tau),\n{\cdot}_{1/\gamma})$ $\forall\gamma\in\,]0,1[$ \cite[Thm. 2.5]{Ayupov:1986} \cite[Cors. 12, 13]{Iochum:1986} is equivalent with uniform Fr\'{e}chet differentiability of $(L_{1/\gamma}(A,\tau),$ $\n{\cdot}_{1/\gamma})$ $\forall\gamma\in\,]0,1[$. Given a polar decomposition $x=s_x\jordan\ab{x}$ with $s_x\in A$ such that $s_x^2=\II$, the formula $\n{x}_{1/\gamma}^{1-1/\gamma}s_x\jordan\ab{x}^{1/\gamma-1}$ \cite[p. 51]{Abdullaev:1984} \cite[Lem. V.3.3.2$^o$]{Iochum:1984} (cf. \cite[p. 101]{Ayupov:1986} and \cite[p. 420]{Iochum:1986}) equals to $\DF\n{x}_{1/\gamma}$ by \cite[Lem. 14]{Iochum:1986}. Hence, using $j(x)=\frac{1}{2}\DF(\n{x}_X^2)=\n{x}_X\DF\n{x}_X$, which is valid for any Fr\'{e}chet differentiable $(X,\n{\cdot}_X)$, we obtain $j(x)=\n{x}_{1/\gamma}^{2-1/\gamma}s_x\jordan\ab{x}^{1/\gamma-1}$. Furthermore, $\gamma$-Lipschitz--H\"{o}lder (resp., Lipschitz) continuity of $\ell_\gamma$ (resp., $(\ell_\gamma)^\inver$), proved in Proposition \ref{prop.na.l.gamma.hoelder}.\ref{prop.na.l.gamma.hoelder.ii} (resp., \ref{prop.na.l.gamma.hoelder}.\ref{prop.na.l.gamma.hoelder.i}), implies (uniform continuity, hence also) norm-to-norm continuity of $\ell_\gamma$ (resp., $(\ell_\gamma)^\inver$) on $(B(A_\star,\n{\cdot}_1))^+$ (resp., $(B(L_{1/\gamma},\n{\cdot}_{1/\gamma}))^+$). The rest of the proof follows from Propositions \ref{prop.D.ell.psi.properties} and \ref{prop.left.pythagorean.info.Psi.alpha.beta}.\ref{prop.left.pythagorean.info.Psi.alpha.beta.ii}--\ref{prop.left.pythagorean.info.Psi.alpha.beta.ix} in the same way as in the Proposition \ref{prop.Wstar.gamma.alpha.beta} and Corollary \ref{cor.d.gamma}.\ref{cor.d.gamma.i}.
\end{proof}

\begin{remark}\label{remark.postquantum.Lp}
Any JBW-algebra (hence, also a self-adjoint part of any W$^*$-algebra) is a special case of an archimedean order unit space $(A,\n{\cdot}_{A})$ with a distinguished order unit $e$, which is Banach dual to the radially compact base normed space $(V,\n{\cdot}_{V}) \iso(A_\star,\n{\cdot}_{A_\star})$. Hence, it is natural to ask whether the above results can be extended to radially compact base normed spaces. If these spaces satisfy an additional spectral duality condition \cite[Def. (p. 55)]{Alfsen:Shultz:1976}, then they admit spectral theory and functional calculus \cite[\S7--\S8]{Alfsen:Shultz:1976}. The notion of a finite trace $\tau_{\mathrm{AS}}$ on such $(A,\n{\cdot}_{A})$ has been introduced in \cite[Def. (p. 107)]{Alfsen:Shultz:1976}, and was extended beyond finite case in \cite[Def. 2.2]{Tikhonov:1993}. Construction of a norm $\n{\cdot}_{1/\gamma}:=(\tau(\ab{\cdot}^{1/\gamma}))^\gamma$ on $A_{1/\gamma,\tau}:=\{x\in A\mid x\geq0, (\tau(x))^\gamma<\infty\}$, implying the construction of Banach spaces $(L_{1/\gamma}(A,\tau):=\overline{A_{1/\gamma,\tau}}^{\n{\cdot}_{1/\gamma}},\n{\cdot}_{1/\gamma})$ with $\gamma\in\,]0,1[$, was provided in \cite[Cor. 3.12]{Tikhonov:1993} for (not necessarily finite) $\tau=\tau_{\mathrm{AS}}$. However, $(L_1(A,\tau_{\mathrm{AS}}),\n{\cdot}_1)\iso(A_\star,\n{\cdot}_{A_\star})$ if{}f $A$ is a JBW-algebra with $e=\II$ \cite[Thm. 6]{Berdikulov:2006}. Another notion of a trace on $A$, $\tau_{\mathrm{B}}$, has been proposed in \cite[Def. 1]{Berdikulov:2005}, together with a corresponding norm $\n{\cdot}_1$ on $A_{1,\tau_{\mathrm{B}}}$ \cite[Thm. 1]{Berdikulov:2005}, and with a proof that $(L_1(A,\tau_{\mathrm{B}}),\n{\cdot}_1)\iso(A_\star,\n{\cdot}_{A_\star})$ for any order unit space $(A,\n{\cdot}_{A})$ in spectral duality \cite[Thm. 2]{Berdikulov:2005}. Hence, in order to generalise our results for any $(V,\n{\cdot}_{V})$ in spectral duality, the following statements have to be proved: (i) $\n{\cdot}_{1/\gamma}$ determined by a faithful $\tau_{\mathrm{B}}$ is a norm on $A_{1/\gamma,\tau_{\mathrm{B}}}$; (ii) $(L_{1/\gamma}(A,\tau_{\mathrm{B}}),\n{\cdot}_{1/\gamma})$ are reflexive, Gateaux differentiable, and strictly convex (cf. Proposition \ref{prop.BBC.lemma}); (iii) they are also Fr\'{e}chet differentiable and have Radon--Riesz--Shmul'yan property (cf. Proposition \ref{prop.Resmerita.cor}); (iv) the corresponding generalisation of $\ell_{\gamma}$ is norm-to-norm continuous (cf. Proposition \ref{prop.D.ell.psi.properties}.\ref{prop.D.ell.psi.properties.v}). Below we make the first steps in this direction.
\end{remark}

\begin{proposition}\label{prop.berdikulov.A.gamma}
Let $(A,\n{\cdot}_A)$ be an archimedean order unit space, which is in spectral duality with a radially compact base normed space $(V,\n{\cdot}_V)\iso(A_\star,\n{\cdot}_{A_\star})$. Let $\tau:A^+\ra\RR^+$ be a finite (resp., finite and faithful) Berdikulov trace, as defined by \textup{\cite[Def. 1]{Berdikulov:2005}}. If $\gamma\in\,]0,1]$, then the function $x\mapsto\n{x}_{1/\gamma}:=(\tau(\ab{x}^{1/\gamma}))^\gamma$ is a seminorm (resp., norm) on $A_{1/\gamma,\tau}:=\{x\in A\mid\n{x}_{1/\gamma}<\infty\}$.
\end{proposition}
\begin{proof}
Follows from \cite[Cor. 3.12]{Tikhonov:1993}, combined with the fact that a finite Berdikulov trace is a finite Alfsen--Shultz trace \cite[Lem. 1]{Berdikulov:2005}.
\end{proof}

\begin{definition}\label{def.L.gamma.order.unit}
Let $(A,\n{\cdot}_A)$ be an archimedean order unit space that is spectrally dual to a radially compact base normed space $(V,\n{\cdot}_V)$. Let $\tau$ be a finite faithful Berdikulov trace on $A$. Let $\gamma,\gamma_1,\gamma_2\in\,]0,1]$. Then:
\begin{enumerate}[nosep,label=(\roman*)]
\item\label{def.L.gamma.order.unit.i} the \df{$L_{1/\gamma}(A,\tau)$ space} is defined as a completion of $A_{1/\gamma,\tau}$ in the norm $\n{\cdot}_{1/\gamma}$. Furthermore, $(L_\infty(A,\tau),\n{\cdot}_\infty):=(A,\n{\cdot}_A)$;
\item\label{def.L.gamma.order.unit.ii} (a positive part of) the \df{postquantum Mazur map} is defined as
\begin{equation}
\ell_{\gamma_1,\gamma_2}:(L_{1/\gamma_1}(A,\tau))^+\ni\phi\mapsto\phi^{\gamma_2/\gamma_1}\in(L_{1/\gamma_2}(A,\tau))^+.
\end{equation}
\end{enumerate}
\end{definition}

\subsection{Noncommutative Orlicz spaces and Kaczmarz maps}

\begin{definition}\label{def.orlicz.terminology}
\begin{enumerate}[nosep,label=(\roman*)]
\item\label{def.orlicz.terminology.i} if $\orlicz:\RR\ra\RR^+$ is even, convex, and $\orlicz(u)=0$ $\iff$ $u=0$, then it will be called an \df{Orlicz} function (cf. \textup{\cite[p. 208]{Orlicz:1932}});
\item\label{def.orlicz.terminology.ii} a \df{Young--Birnbaum--Orlicz dual} of an Orlicz function $\orlicz$ is defined as  \textup{\cite[p. 226]{Young:1912}} \textup{\cite[Eqn. (5)]{Birnbaum:Orlicz:1931}}
\begin{equation}
\RR\ni y\mapsto\orlicz^\young(y):=\sup\{x\ab{y}-\orlicz(x)\mid x\geq0\}\in[0,\infty];
\end{equation}
\item\label{def.orlicz.terminology.iii}for any Orlicz function $\Orlicz$ and any interval $I\subseteq\RR$, we will denote:  
\begin{align}
\hspace{-1.2cm}
\orlicz'_+\mbox{ (resp., }\orlicz'_+\mbox{)}\;&:=\mbox{a right (resp., left) derivative of }\orlicz,\\
\varpi_\orlicz(\lambda)\;&:=\sup\{t>0\mid\orlicz^\young(\orlicz'_+(t))\leq\lambda\},\\
\orlicz\in\NFUN\;&:\;\iff\;\lim_{u\ra^+0}\textstyle\frac{\orlicz(u)}{u}=0\mbox{ and }\lim_{u\ra\infty}\textstyle\frac{\orlicz(u)}{u}=\infty\nonumber\\&\;\;\;\;\;\;\;\;\;\;\;\;\;\;\textup{\cite[Def. I.\S1.5]{Birnbaum:Orlicz:1931}},\\
\orlicz\in\triangle_2^0\;&:\;\iff\;\lim_{u\ra^+0}\textstyle\frac{\orlicz(2u)}{\orlicz(u)}<\infty\;\;\textup{\cite[Eqn. ($\triangle_2$)]{Birnbaum:Orlicz:1931}},\\
\orlicz\in\triangle_2^\infty\;&:\;\iff\;\limsup_{u\ra\infty}\textstyle\frac{\orlicz(2u)}{\orlicz(u)}<\infty\;\;\textup{\cite[p. 36]{Birnbaum:Orlicz:1931}},\\
\orlicz\in\triangle_2\;&:\;\iff\;\sup_{u>0}\textstyle\frac{\orlicz(2u)}{\orlicz(u)}<\infty\;\;\textup{\cite[p. 494]{Burkill:1928}}\nonumber\\&\;\;\iff\;\orlicz\in\triangle_2^0\cap\triangle_2^\infty,\\
\orlicz\in\SC(I)\;&:\;\iff\;\orlicz\mbox{ is strictly convex on }I,\\
\orlicz\in\DIFF(I)\;&:\;\iff\;\orlicz\mbox{ is continuously differentiable on }I.
\end{align}
\end{enumerate}
\end{definition}

\begin{definition}\label{def.orlicz.spaces}
Let $\N$ be a semifinite W$^*$-algebra, let $\tau$ be a faithful normal semifinite trace on $\N$. Let $\MMM(\N,\tau)$ denote the space of all $\tau$-measurable operators affiliated with $\N$ \textup{\cite[\S2]{Nelson:1974}} \textup{\cite[p. 91]{Yeadon:1975}}. Then:
\begin{enumerate}[nosep,label=(\roman*)]
\item\label{def.orlicz.spaces.i}a \df{noncommutative Orlicz space} is defined as\footnote{For commutative $\N=L_\infty(\X,\mu)$, $L_\orlicz(\X,\mu)$ was introduced in \cite[Prop. 1]{Orlicz:1932}, $\n{\cdot}_\orlicz^{\mathrm{O}}$ was introduced in \cite[p. 210]{Orlicz:1932}, while $\n{\cdot}_\orlicz$ was introduced independently in \cite[Def. (p. 601)]{Morse:Transue:1950} and \cite[p. 181]{Nakano:1950} (this norm is often called the ``Luxemburg norm'', in reference to its appearance in \cite[pp. 43--44]{Luxemburg:1955}, however the latter work explicitly cites Nakano's book \cite{Nakano:1950}). For noncommutative $\N$, $(L_\orlicz(\N,\tau),\n{\cdot}_\orlicz)$ has been introduced in: \textup{\cite[\S2]{Rao:1971}} for type I W$^*$-algebras; \textup{\cite[p. 6]{Muratov:1979}} (=\textup{\cite[Def. 2.3.19, p. 111]{Muratov:1979:PhD}}) for type II$_1$ W$^*$-algebras (with a slightly different, yet isometrically isomorphic, definition of a norm; $\n{\cdot}_\orlicz$ for this case appeared in \cite[\S1]{Chilin:1986}); \cite[Props. 2.1, 2.2]{Bikchentaev:1987} and \cite[p. 126]{Kunze:1990} for type II$_\infty$ W$^*$-algebras. For any semifinite $\N$, $\n{\cdot}_\orlicz^{\mathrm{O}}$ on $L_\orlicz(\N,\tau)$ has been introduced in \cite[p. 127]{Kunze:1990} (for type II$_1$ $\N$, a slightly different, yet isometrically isomorphic, definition appeared earlier in \cite[p. 12]{Muratov:1978}).}
\begin{equation}
L_\orlicz(\N,\tau):=\{x\in\MMM(\N,\tau)\mid\exists\lambda>0\;\tau(\orlicz(\lambda x))<\infty\};
\end{equation}
a \df{noncommutative Morse--Transue--Nakano norm} on $L_\orlicz(\N,\tau)$ is defined as
\begin{equation}
\n{x}_\orlicz:=\inf\{\lambda\geq0\mid\tau(\orlicz(x/\lambda))\leq1\}\;\forall x\in L_\orlicz(\N,\tau);
\end{equation}
a \df{noncommutative Orlicz norm} on  $L_\orlicz(\N,\tau)$ is defined as
\begin{equation}
\hspace{-0.73cm}\n{x}_\orlicz^{\mathrm{O}}:=\sup\{\tau(\ab{xy})\mid y\in\MMM(\N,\tau),\;\tau(\orlicz^\young(\ab{y})\leq1\}\;\forall x\in L_\orlicz(\N,\tau);
\end{equation}
\item\label{def.orlicz.spaces.ii} if $\orlicz_1$ and $\orlicz_2$ are Orlicz functions, then we define a \df{noncommutative Kaczmarz map} as
\begin{equation}
\hspace{-0.4cm}\ell_{\orlicz_1,\orlicz_2}:L_{\orlicz_1}(\N,\tau)\ni x=u_x\ab{x}\mapsto u_x((\orlicz_2)^{\inver}\circ\orlicz_1)(\ab{x})\in L_{\orlicz_2}(\N,\tau),
\label{eqn.noncomm.kaczmarz}
\end{equation}
where $x=u_x\ab{x}$ is the unique polar decomposition of $x$;
\item\label{def.orlicz.spaces.iii}we will say that $\N$ is of \df{type \Isfi}\ if{}f it is a separable factor of type I$_\infty$;
\item\label{def.orlicz.spaces.iv} if $\N$ is either commutative, or of type \Isfi, or of type II$_1$, or of type II$_\infty$, then
\[
\hspace{-0.85cm}\type(\N):=
\left\{
\begin{array}{ll}
\mbox{\Isfi}&\st\N\mbox{ is noncommutative and of type \Isfi},\\
&\,\;\;\mbox{or }\N=L_\infty(\X,\mu)\mbox{ with purely atomic, infinite }(\X,\mu)\\
\mbox{II}_1&\st\N\mbox{ is noncommutative and of type II}_1,\\
&\,\;\;\mbox{or }\N=L_\infty(\X,\mu)\mbox{ with nonatomic, finite }(\X,\mu)\\
\mbox{II}_\infty&\st\N\mbox{ is noncommutative and of type II}_\infty,\\&\,\;\;\mbox{or }
\N=L_\infty(\X,\mu)\mbox{ with nonatomic, infinite }(\X,\mu).
\end{array}
\right.
\]
\end{enumerate}
\end{definition}

\begin{proposition}\label{prop.nc.Orlicz.spaces.Banach.duality}
Let $\N$ be a W$^*$-algebra either of type \Isfi, or of type II$_1$, or of type II$_\infty$, let $\tau$ be a faithful normal semifinite trace on $\N$. Let $\orlicz$ be an Orlicz function such that $\orlicz^\young$ is an Orlicz function. Let $\orlicz\in\triangle_2^0$ (resp., $\triangle_2^\infty$; $\triangle_2$) if $\N$ is of type \Isfi (resp., II$_1$; II$_\infty$). Then
\begin{equation}
(L_\orlicz(\N,\tau),\n{\cdot}_\orlicz)^\star\iso(L_{\orlicz^\young}(\N,\tau),\n{\cdot}_{\orlicz^\young}^{\mathrm{O}}).
\label{eqn.nc.Orlicz.spaces.Banach.duality}
\end{equation}
\end{proposition}
\begin{proof}
The proof is based on mutual relationship of commutative \cite[p. 1292]{Semenov:1964} \cite[II.\S4.1]{Krein:Petunin:Semenov:1978} and noncommutative \cite[3$^o$]{Ovchinnikov:1970} \cite[p. 10]{Medzhitov:1987} rearrangement invariant spaces, denoted $(E(\X,\mu),\n{\cdot}_{E(\X,\mu)})$ and $(E(\N,\tau),\n{\cdot}_{E(\N,\tau)})$, respectively, with $\type(\N)=\type(L_\infty(\X,\mu))$,
\begin{equation}
\;E(\N,\tau):=\{x\in\MMM(\N,\tau)\mid x^\tau\in E(\X,\mu)\},\;\mbox{and}\;\n{\cdot}_{E(\N,\tau)}:=\n{(\cdot)^\tau}_{E(\X,\mu)},
\label{eqn.riBs.ncriBs}
\end{equation}
\sloppy where $x^\tau$ denotes a rearrangement of $x\in E(\N,\tau)$ \cite[\S4]{Grothendieck:1955} \cite[p. 79]{Ovchinnikov:1970:snumbers} \cite[Def. 2.2]{Yeadon:1975}. If $(E(\X,\mu),\n{\cdot}_{E(\X,\mu)})$ is order continuous, then it is strongly symmetric \cite[Cor. (p. 142)]{Krein:Petunin:Semenov:1978}. If $(E(\X,\mu),$ $\n{\cdot}_{E(\X,\mu)})$ is strongly symmetric, then $(E(\N,\tau),\n{\cdot}_{E(\N,\tau)})$ is strongly symmetric \cite[Thm. 51]{Dodds:dePagter:2014} (=\cite[Thm. 6.1.2, Prop. 6.8.13.(i)]{Dodds:dePagter:Sukochev:2023}). If $(E(\N,\tau),\n{\cdot}_{E(\N,\tau)})$ is strongly symmetric and $\N$ is of type \Isfi, II$_1$, or II$_\infty$, then its noncommutative K\"{o}the--Toeplitz\footnote{\cite[Defs. \S2.1, \S2.2]{Koethe:Toeplitz:1934} introduced this duality for atomic infinite $(\X,\mu)$, while \cite[pp. 99--100]{Dieudonne:1952} and \cite[\S2]{Cooper:1953}, independently, have extended it to nonatomic $(\X,\mu)$. There is no valid reason to exclude Toeplitz from terminology, since, like K\"{o}the, he also has published an independent research on this topic after their joint paper \cite{Koethe:Toeplitz:1934}, and both of them have considered only the case of atomic infinite $(\X,\mu)$. Cf. also \cite[\S3.5]{Lorentz:1953} and \cite[p. 9]{Luxemburg:1955} for the same terminological choice.} dual $((E(\N,\tau))^\koethe,(\n{\cdot}_{E(\N,\tau)})^\koethe)$ \cite[p. 227]{Dodds:dePagter:2014} satisfies $((E(\N,\tau))^\koethe,(\n{\cdot}_{E(\N,\tau)})^\koethe)=(E(\N,\tau),\n{\cdot}_{E(\N,\tau)})^\star$ \cite[Thm. 27]{Dodds:dePagter:2014}. For any Orlicz $\orlicz$, if $\orlicz\in\triangle_2^0$ (resp., $\triangle_2^\infty$; $\triangle_2$) for $\type(L_\infty(\X,\mu))=$ \Isfi{} (resp., II$_1$; II$_\infty$), then $(L_\orlicz(\X,\mu),\n{\cdot}_\orlicz)$ is order continuous \cite[Thm. 2.3.6]{Luxemburg:1955} and $(L_\orlicz(\X,\mu),\n{\cdot}_{\orlicz})^\star\iso(L_{\orlicz^\young}(\X,\mu),\n{\cdot}_{\orlicz^\young}^{\mathrm{O}})$ \cite[Cor. 2.2.10, Thms. 2.2.11, 2.1.17]{Edgar:Sucheston:1992}. Hence, $(L_\orlicz(\N,\tau),\n{\cdot}_\orlicz)^\star\iso(L_{\orlicz^\young}(\N,\tau),(\n{\cdot}_\orlicz)^\koethe=\n{(\cdot)^\tau}_{\orlicz^\young}^{\mathrm{O}})$. Finally, from $\n{x}_\orlicz\leq1$ $\iff$ $\tau(\orlicz(\ab{x}))\leq 1$ $\forall x\in L_\orlicz(\N,\tau)$ \cite[Lem. 5.40]{Goldstein:Labuschagne:2020} and a definition of $(\n{\cdot}_{E(\N,\tau)})^\koethe$ \cite[p. 227]{Dodds:dePagter:2014} it follows that $\n{\cdot}_{\orlicz^\young}^{\mathrm{O}}=(\n{\cdot}_{\orlicz})^\koethe$.
\end{proof}

\begin{proposition}\label{prop.reflexivity.riBs.type.Isf.infty}
Let $(E(\N,\tau),\n{\cdot}_{E(\N,\tau)})$ and $(E(\X,\mu),\n{\cdot}_{E(\X,\mu)})$ be rearrangement invariant Banach spaces, mutually related by \eqref{eqn.riBs.ncriBs}. Let $\N$ be of type \Isfi. If $(E(\N,\tau),\n{\cdot}_{E(\N,\tau)})$ is strongly symmetric and reflexive, then $(E(\X,\mu),\n{\cdot}_{E(\X,\mu)})$ is reflexive.
\end{proposition}
\begin{proof}
By \cite[Thm. 45]{Dodds:dePagter:2014}, if $(E(\N,\tau),\n{\cdot}_{E(\N,\tau)})$ is strongly symmetric, then it is reflexive if{}f (it has the Fatou property, and both $(E(\N,\tau),\n{\cdot}_{E(\N,\tau)})$ and $(E(\N,\tau),\n{\cdot}_{E(\N,\tau)})^\koethe$ are order continuous). By \cite[Prop. 5.3]{Medzhitov:1988} (=\cite[Prop. 2]{Medzhitov:1987:atom}) (resp., \cite[Cor. 5.5.1]{Medzhitov:1988} (=\cite[Cor. 2.1]{Medzhitov:1987:atom})), $(E(\N,\tau),\n{\cdot}_{E(\N,\tau)})$ is order continuous (resp., satisfies the Fatou property) if{}f $(E(\X,\mu),\n{\cdot}_{E(\X,\mu)})$ is order continuous (resp., satisfies the Fatou property). Furthermore, by \cite[Lemm. 5.4]{Medzhitov:1988} (=\cite[Lem. (p. 71)]{Medzhitov:1987:atom}), the sequence space corresponding to $(E(\N,\tau))^\koethe$ coincides with $(E(\X,\mu))^\koethe$. Finally, by \cite[Thm. \S3.1]{Ogasawara:1944}, $(E(\X,\mu),\n{\cdot}_{E(\X,\mu)})$ is reflexive if{}f it has the Fatou property and the spaces $(E(\X,\mu),\n{\cdot}_{E(\X,\mu)})$ and $(E(\X,\mu),\n{\cdot}_{E(\X,\mu)})^\koethe$ are order continuous.
\end{proof}

\begin{corollary}\label{cor.reflexivity.riBs.type.Isf.infty}
Let $(E(\N,\tau),\n{\cdot}_{E(\N,\tau)})$ and $(E(\X,\mu),\n{\cdot}_{E(\X,\mu)})$ be rearrangement invariant Banach spaces, mutually related by \eqref{eqn.riBs.ncriBs}. Let $\N$ be of type \Isfi, and let either $(E(\N,\tau),\n{\cdot}_{E(\N,\tau)})$ or $(E(\X,\mu),\n{\cdot}_{E(\X,\mu)})$ be order continuous. Then $(E(\N,\tau),\n{\cdot}_{E(\N,\tau)})$ is reflexive if{}f $(E(\X,\mu),\n{\cdot}_{E(\X,\mu)})$ is reflexive.
\end{corollary}
\begin{proof}
By \cite[Prop. 6.8.15]{Dodds:dePagter:Sukochev:2023} (cf. also \cite[p. 153]{Arazy:1981:ball}), if $\N$ is of type \Isfi, and $(E(\X,\mu),\n{\cdot}_{E(\X,\mu)})$ is reflexive, then $(E(\N,\tau),\n{\cdot}_{E(\N,\tau)})$ is reflexive. The conclusion follows from taking into account Proposition \ref{prop.reflexivity.riBs.type.Isf.infty}, equivalence of order continuity of $(E(\N,\tau),\n{\cdot}_{E(\N,\tau)})$ and $(E(\X,\mu),\n{\cdot}_{E(\X,\mu)})$ \cite[Prop. 5.3]{Medzhitov:1988} (=\cite[Prop. 2]{Medzhitov:1987:atom}), and the fact that order continuity of $(E(\N,\tau),\n{\cdot}_{E(\N,\tau)})$ implies its strong symmetry \cite[Cor. 5.3.4]{Dodds:dePagter:Sukochev:2023}.
\end{proof}

\begin{proposition}
\label{prop.noncommutative.MTNL.norm.geometry.characterisation}
Let $\N$ be a W$^*$-algebra either of type \Isfi{}, or of type II$_1$, or of type II$_\infty$, let $\tau$ be a faithful normal semifinite trace on $\N$, let $(\X,\mu)$ be a countably finite measure space such that $\type(L_\infty(\X,\mu))=\type(\N)$. Let $\orlicz:\RR\ra\RR^+$ be an Orlicz function. Then:
\begin{enumerate}[nosep,label=(\roman*)]
\item\label{prop.noncommutative.MTNL.norm.geometry.characterisation.i}equivalent are:
\begin{enumerate}[nosep,label=\alph*)]
\item\label{prop.noncommutative.MTNL.norm.geometry.characterisation.i.1} $(L_\orlicz(\N,\tau),\n{\cdot}_{\orlicz})$ is strictly convex;
\item\label{prop.noncommutative.MTNL.norm.geometry.characterisation.i.2} $(L_\orlicz(\X,\mu),\n{\cdot}_{\orlicz})$ is strictly convex;
\item\label{prop.noncommutative.MTNL.norm.geometry.characterisation.i.3}$%
\left\{
\begin{array}{ll}
\orlicz\in\triangle_2^0\cap\SC([0,\orlicz^{\inver}(\frac{1}{2})])
&\st\type(\N)=\mbox{\Isfi}\\
\orlicz\in\triangle_2^\infty\cap\SC(\RR)
&\st\type(\N)=\mbox{II}_1\\
\orlicz\in\triangle_2\cap\SC(\RR)
&\st\type(\N)=\mbox{II}_\infty;
\end{array}
\right.$
\end{enumerate}
\item\label{prop.noncommutative.MTNL.norm.geometry.characterisation.ii} if $\orlicz^\young\in\triangle_2^\infty$ (resp., $\triangle_2$) for $\type(\N)$ $=$ II$_1$ (resp., II$_\infty$), then equivalent are:
\begin{enumerate}[nosep,label=\alph*)]
\item\label{prop.noncommutative.MTNL.norm.geometry.characterisation.ii.1} $(L_\orlicz(\N,\tau),\n{\cdot}_{\orlicz})$ is Gateaux differentiable;
\item\label{prop.noncommutative.MTNL.norm.geometry.characterisation.ii.2}$(L_\orlicz(\X,\mu),\n{\cdot}_{\orlicz})$ is Gateaux differentiable;
\item\label{prop.noncommutative.MTNL.norm.geometry.characterisation.ii.3} $%
\left\{
\begin{array}{ll}
\orlicz\in\triangle_2^0\cap\DIFF([0,\orlicz^{\inver}(1)[)
&\st\type(\N)=\mbox{\Isfi}\\
\orlicz\in\triangle_2^\infty\cap\DIFF(\RR)
&\st\type(\N)=\mbox{II}_1\\
\orlicz\in\triangle_2\cap\DIFF(\RR)
&\st\type(\N)=\mbox{II}_\infty;
\end{array}
\right.$
\end{enumerate}
\item\label{prop.noncommutative.MTNL.norm.geometry.characterisation.iii} if $\orlicz\in\NFUN$ for $\type(\N)=\mbox{II}_1$, then equivalent are:
\begin{enumerate}[nosep,label=\alph*)]
\item\label{prop.noncommutative.MTNL.norm.geometry.characterisation.iii.1} $(L_\orlicz(\N,\tau),\n{\cdot}_{\orlicz})$ has Radon--Riesz--Shmul'yan property;
\item\label{prop.noncommutative.MTNL.norm.geometry.characterisation.iii.2}$(L_\orlicz(\X,\mu),\n{\cdot}_{\orlicz})$ has Radon--Riesz--Shmul'yan property;
\item\label{prop.noncommutative.MTNL.norm.geometry.characterisation.iii.3}$%
\left\{
\begin{array}{ll}
\orlicz\in\triangle_2^0
&\st\type(\N)=\mbox{\Isfi}\\
\orlicz\in\triangle_2^\infty\cap\SC(\RR)
&\st\type(\N)=\mbox{II}_1\\
\orlicz\in\triangle_2\cap\SC(\RR)
&\st\type(\N)=\mbox{II}_\infty;
\end{array}
\right.$
\end{enumerate}
\item\label{prop.noncommutative.MTNL.norm.geometry.characterisation.iv} if $\type(\N)=\mbox{II}_1$ or ($\orlicz\in\triangle_2$ if $\type(\N)=\mbox{II}_\infty$) then equivalent are:
\begin{enumerate}[nosep,label=\alph*)]
\item\label{prop.noncommutative.MTNL.norm.geometry.characterisation.iv.1} $(L_\orlicz(\N,\tau),\n{\cdot}_{\orlicz})$ is reflexive;
\item\label{prop.noncommutative.MTNL.norm.geometry.characterisation.iv.2} $(L_\orlicz(\X,\mu),\n{\cdot}_{\orlicz})$ is reflexive;
\item\label{prop.noncommutative.MTNL.norm.geometry.characterisation.iv.3}$%
\left\{
\begin{array}{ll}
\orlicz,\orlicz^\young\in\triangle_2^0
&\st\type(\N)=\mbox{\Isfi}\\
\orlicz,\orlicz^\young\in\triangle_2^\infty
&\st\type(\N)=\mbox{II}_1\\
\orlicz^\young\in\triangle_2
&\st\type(\N)=\mbox{II}_\infty.
\end{array}
\right.$
\end{enumerate}
\end{enumerate}
\end{proposition}
\begin{proof}
In what follows, we restrict the mutual characterisations of  geometric properties of norms of corresponding commutative and noncommutative rearrangement invariant Banach spaces (denoted, respectively, $(E(\X,\mu),\n{\cdot}_{E(\X,\mu)})$ and $(E(\N,\tau),\n{\cdot}_{E(\N,\tau)})$, and mutually related via \eqref{eqn.riBs.ncriBs}) to the case of noncommutative Orlicz spaces, and combine them with the corresponding characterisations of geometric properties of norms on commutative Orlicz spaces by means of the properties of Orlicz function.
\begin{enumerate}[nosep]
\item[(i)] \ref{prop.noncommutative.MTNL.norm.geometry.characterisation.i.3} $\iff$ \ref{prop.noncommutative.MTNL.norm.geometry.characterisation.i.2} is proved in: \cite[Rem. 2]{Kaminska:1981} for $\type(\N)=\mbox{\Isfi}$; \cite[Cor. 5]{Turett:1976} for $\type(\N)=\mbox{II}_1$; \cite[Thm. 1.7]{Fennich:1980} for $\type(\N)=\mbox{II}_\infty$. \ref{prop.noncommutative.MTNL.norm.geometry.characterisation.i.1} $\iff$ \ref{prop.noncommutative.MTNL.norm.geometry.characterisation.i.2} follows as a special case of: \cite[Cor. 2.5.(i)]{Arazy:1981:ball} for $(E(\N,\tau),\n{\cdot}_{E(\N,\tau)})$ with $\type(\N)=\mbox{\Isfi}$; \cite[Thm. 1.1]{Chilin:Krygin:Sukochev:1992:extreme} \cite[Cor. 5.6]{Czerwinska:Kaminska:2017} for $(E(\N,\tau),\n{\cdot}_{E(\N,\tau)})$ with $\type(\N)\in\{\mbox{II}_1,\mbox{II}_\infty\}$.
\item[(ii)] \ref{prop.noncommutative.MTNL.norm.geometry.characterisation.ii.3} $\iff$ \ref{prop.noncommutative.MTNL.norm.geometry.characterisation.ii.2} is proved in: \cite[Thm. 13]{Grzaslewicz:Hudzik:1992} for $\type(\N)=\mbox{\Isfi}$; \cite[Thm. 11]{Grzaslewicz:Hudzik:1992} for $\type(\N)\in\{\mbox{II}_1,\mbox{II}_\infty\}$. \ref{prop.noncommutative.MTNL.norm.geometry.characterisation.ii.1} $\iff$ \ref{prop.noncommutative.MTNL.norm.geometry.characterisation.ii.2} follows as a special case of: \cite[Cor. 2.5.(ii)]{Arazy:1981:ball} for $(E(\N,\tau),\n{\cdot}_{E(\N,\tau)})$ with $\type(\N)=\mbox{\Isfi}$; \cite[Cor. 2.13]{Czerwinska:Kaminska:Kubiak:2012} for $(E(\N,\tau),\n{\cdot}_{E(\N,\tau)})$ with $\type(\N)\in\{\mbox{II}_1,\mbox{II}_\infty\}$. Both \cite[Cor. 2.5.(ii)]{Arazy:1981:ball} and \cite[Cor. 2.13]{Czerwinska:Kaminska:Kubiak:2012} require order continuity of $(E(\X,\mu),\n{\cdot}_{E(\X,\mu)})$. For $(L_\orlicz(\X,\mu),\n{\cdot}_\orlicz)$ this is imposed by $\orlicz\in\triangle_2^0$ (resp., $\triangle_2^\infty$; $\triangle_2$) for $\type(\N)$ = \Isfi (resp., II$_1$; II$_\infty$). \cite[Cor. 2.13]{Czerwinska:Kaminska:Kubiak:2012} requires also $\lim_{t\ra\infty}\rearr{x}{\mu}(t)=0$ $\forall x\in(E(\X,\mu))^\koethe$, which is satisfied if $((E(\X,\mu))^\koethe,(\n{\cdot}_{E(\X,\mu)})^\koethe)$ is order continuous (cf., e.g., \cite[p. 730]{Dodds:Dodds:dePagter:1993}).
\item[(iii)] \ref{prop.noncommutative.MTNL.norm.geometry.characterisation.iii.3} $\iff$ \ref{prop.noncommutative.MTNL.norm.geometry.characterisation.iii.2} is proved in: \cite[Thm. 2.8]{Hudzik:Pallaschke:1997} for $\type(\N)=\mbox{\Isfi}$; \cite[Thm. (p. 61)]{Chen:Wang:1987} and \cite[Thm. 3]{Medzhitov:Sukochev:1992} $\type(\N)=\mbox{II}_1$; \cite[Thm. (p. 341)]{Wang:Cui:Zhang:1998} for $\type(\N)=\mbox{II}_\infty$. \ref{prop.noncommutative.MTNL.norm.geometry.characterisation.iii.1} $\iff$ \ref{prop.noncommutative.MTNL.norm.geometry.characterisation.iii.2} follows as a special case of: \cite[Thm. I]{Arazy:1981:more} for $(E(\N,\tau),\n{\cdot}_{E(\N,\tau)})$ with $\type(\N)=\mbox{\Isfi}$; \cite[Thm. 2.7]{Chilin:Dodds:Sukochev:1997} (cf. \cite[Thm. 16.4]{Czerwinska:Kaminska:2017}) for $(E(\N,\tau),\n{\cdot}_{E(\N,\tau)})$ with $\type(\N)\in\{\mbox{II}_1,\mbox{II}_\infty\}$.
\item[(iv)]  \ref{prop.noncommutative.MTNL.norm.geometry.characterisation.iv.3} $\iff$ \ref{prop.noncommutative.MTNL.norm.geometry.characterisation.iv.2} is proved in \cite[Thm. 5]{Luxemburg:1955} for $\type(\N)\in\{\mbox{\Isfi},\mbox{II}_1,\mbox{II}_\infty\}$. If $\type(\N)=\mbox{\Isfi}$ (resp., II$_1$; II$_\infty$), then \ref{prop.noncommutative.MTNL.norm.geometry.characterisation.iv.1} $\iff$ \ref{prop.noncommutative.MTNL.norm.geometry.characterisation.iv.2} follows as a special case of Corollary \ref{cor.reflexivity.riBs.type.Isf.infty} (resp., \cite[Thm. 1.3.6]{Sukochev:1987}; \cite[Thm. 4.8]{Dodds:Dodds:1995}). \cite[Thm. 4.8]{Dodds:Dodds:1995} requires $(E(\X,\mu),\n{\cdot}_{E(\X,\mu)})$ to be strongly symmetric. This is implied by the order continuity of  $(E(\X,\mu),\n{\cdot}_{E(\X,\mu)})$ \cite[Cor. (p. 142)]{Krein:Petunin:Semenov:1978}. The latter is implied for $(L_\orlicz(\X,\mu),\n{\cdot}_\orlicz)$ for $\type(\N)$ $=$ II$_\infty$ by $\orlicz\in\triangle_2$ due to \cite[Thm. 2.3.6]{Luxemburg:1955}.
\end{enumerate}
\end{proof}

\begin{remark}
For Orlicz $\orlicz$ and $\orlicz^\young$, the Banach duality \eqref{eqn.nc.Orlicz.spaces.Banach.duality} appeared in \cite[Thm. 2.4]{Ma:Jiang:Ji:2018} (=\cite[Thm. 3.3.1]{Ma:2017}), under condition $\orlicz\in\triangle_2^\infty$, and without a restriction on a type of $\N$. Proposition \ref{prop.nc.Orlicz.spaces.Banach.duality} shows that this is false if{}f $\type(\N)$ $\neq$ II$_1$. The same conclusion, following from Proposition \ref{prop.noncommutative.MTNL.norm.geometry.characterisation}.\ref{prop.noncommutative.MTNL.norm.geometry.characterisation.iv}, applies to characterisation of reflexivity of $(L_\orlicz(\N,\tau),\n{\cdot}_\orlicz)$ stated in \cite[Cor. 2.3]{Ma:Jiang:Ji:2018} (=\cite[Cor. 3.3.2]{Ma:2017}). Hence, the only novelty contained in the latter corollary, as compared with the characterisation of reflexivity of $(L_\orlicz(\N,\tau),\n{\cdot}_\orlicz)$ for type II$_1$ provided earlier in \cite[Thm. 2]{Muratov:1979} (cf. \cite[2.4.8.(i)]{Muratov:1979:PhD}) and \cite[Cor. 2.9]{Kunze:1990}, amounts to dropping the assumptions of $\orlicz,\orlicz^\young\in\NFUN$. The special case of Proposition \ref{prop.noncommutative.MTNL.norm.geometry.characterisation}.\ref{prop.noncommutative.MTNL.norm.geometry.characterisation.i}, for $\orlicz\in\NFUN$ and $\type(\N)=\mbox{\Isfi}$, was obtained in \cite[Thm. 5.3.3]{Ma:2017}, although with a mistake, using $\orlicz\in\triangle_2$ instead of $\orlicz\in\triangle_2^0$.
\end{remark}

\begin{proposition}
\label{prop.noncommutative.Orlicz.norm.geometry.characterisation}
Let $\N$ be a W$^*$-algebra either of type \Isfi{}, or of type II$_1$, or of type II$_\infty$, let $\tau$ be a faithful normal semifinite trace on $\N$, let $(\X,\mu)$ be a countably finite measure space such that $\type(L_\infty(\X,\mu))=\type(\N)$. Let $\orlicz:\RR\ra\RR^+$ be an Orlicz function. Then:
\begin{enumerate}[nosep,label=(\roman*)]
\item\label{prop.noncommutative.Orlicz.norm.geometry.characterisation.i} equivalent are:
\begin{enumerate}[nosep,label=\alph*)]
\item\label{prop.noncommutative.Orlicz.norm.geometry.characterisation.i.1} $(L_\orlicz(\N,\tau),\n{\cdot}_{\orlicz}^{\mathrm{O}})$ is strictly convex;
\item\label{prop.noncommutative.Orlicz.norm.geometry.characterisation.i.2}$(L_\orlicz(\X,\mu),\n{\cdot}_{\orlicz}^{\mathrm{O}})$ is strictly convex;
\item\label{prop.noncommutative.Orlicz.norm.geometry.characterisation.i.3} $%
\left\{
\begin{array}{ll}
\orlicz\in\SC([0,\varpi_\orlicz(1)]),\;\exists u>0\;\;\orlicz^\young(\orlicz_+'(u))\geq\frac{1}{2}\\
\;\;\;\;\;\;\;\;\;\st\type(\N)=\mbox{\Isfi}\\
\orlicz\in\SC(\RR),\;\lim_{u\ra\infty}((\lim_{t\ra\infty}\frac{\orlicz(t)}{t})\ab{u}-\orlicz(u))=\infty
\\
\;\;\;\;\;\;\;\;\;\st\type(\N)\in\{\mbox{II}_1,\mbox{II}_\infty\};
\end{array}
\right.$
\end{enumerate}
\item\label{prop.noncommutative.Orlicz.norm.geometry.characterisation.ii} if $\lim_{t\ra\infty}\frac{\orlicz(t)}{t}=\infty$, then equivalent are:
\begin{enumerate}[nosep,label=\alph*)]
\item\label{prop.noncommutative.Orlicz.norm.geometry.characterisation.ii.1} $(L_\orlicz(\N,\tau),\n{\cdot}_{\orlicz}^{\mathrm{O}})$ is Gateaux differentiable;
\item\label{prop.noncommutative.Orlicz.norm.geometry.characterisation.ii.2} $(L_\orlicz(\X,\mu),\n{\cdot}_{\orlicz}^{\mathrm{O}})$ is Gateaux differentiable;
\item\label{prop.noncommutative.Orlicz.norm.geometry.characterisation.ii.3} $%
\left\{
\begin{array}{ll}
\orlicz\in\triangle_2^0\cap\DIFF([0,\varpi_\orlicz(\frac{1}{2})[),&\\
\;\;\;\;\;\;\orlicz^\young(\orlicz'_-(\varpi_\orlicz(\frac{1}{2})))=\frac{1}{2}
&\st\type(\N)=\mbox{\Isfi}\\
\orlicz\in\triangle_2^\infty\cap\DIFF(\RR)
&\st\type(\N)=\mbox{II}_1\\
\orlicz\in\triangle_2\cap\DIFF(\RR)
&\st\type(\N)=\mbox{II}_\infty;
\end{array}
\right.$
\end{enumerate}
\item\label{prop.noncommutative.Orlicz.norm.geometry.characterisation.iii} if $\type(\N)\neq\mbox{II}_\infty$ and ($\lim_{u\ra\infty}\frac{\orlicz(u)}{u}=\infty$ if $\type(\N)=\mbox{II}_1$) then equivalent are:
\begin{enumerate}[nosep,label=\alph*)]
\item\label{prop.noncommutative.Orlicz.norm.geometry.characterisation.iii.1} $(L_\orlicz(\N,\tau),\n{\cdot}_{\orlicz}^{\mathrm{O}})$ has Radon--Riesz--Shmul'yan property;
\item\label{prop.noncommutative.Orlicz.norm.geometry.characterisation.iii.2} $(L_\orlicz(\X,\mu),\n{\cdot}_{\orlicz}^{\mathrm{O}})$ has Radon--Riesz--Shmul'yan property;
\item\label{prop.noncommutative.Orlicz.norm.geometry.characterisation.iii.3} $%
\left\{
\begin{array}{ll}
\orlicz\in\triangle_2^0
&\st\type(\N)=\mbox{\Isfi}\\
\orlicz\in\triangle_2^\infty\cap\SC(\RR)
&\st\type(\N)=\mbox{II}_1;
\end{array}
\right.$
\end{enumerate}
\item\label{prop.noncommutative.Orlicz.norm.geometry.characterisation.iv} if $\type(\N)=\mbox{II}_1$ or ($\orlicz\in\triangle_2$ if $\type(\N)=\mbox{II}_\infty$) then equivalent are:
\begin{enumerate}[nosep,label=\alph*)]
\item\label{prop.noncommutative.Orlicz.norm.geometry.characterisation.iv.1} $(L_\orlicz(\N,\tau),\n{\cdot}_{\orlicz}^{\mathrm{O}})$ is reflexive;
\item\label{prop.noncommutative.Orlicz.norm.geometry.characterisation.iv.2} $(L_\orlicz(\X,\mu),\n{\cdot}_{\orlicz}^{\mathrm{O}})$ is reflexive;
\item\label{prop.noncommutative.Orlicz.norm.geometry.characterisation.iv.3} $%
\left\{
\begin{array}{ll}
\orlicz,\orlicz^\young\in\triangle_2^0
&\st\type(\N)=\mbox{\Isfi}\\
\orlicz,\orlicz^\young\in\triangle_2^\infty
&\st\type(\N)=\mbox{II}_1\\
\orlicz^\young\in\triangle_2
&\st\type(\N)=\mbox{II}_\infty.
\end{array}
\right.$
\end{enumerate}
\end{enumerate}
\end{proposition}
\begin{proof}
The method of proof is completely analogous to the method used in the proof of Proposition \ref{prop.noncommutative.MTNL.norm.geometry.characterisation}, so we will only specify references characterising the corresponding properties of commutative Orlicz spaces with Orlicz norm:
\begin{enumerate}[nosep]
\item[(i)] $\type(\N)$ $=$ \Isfi\ (resp., II): \cite[Thm. 2.16]{Cui:Hudzik:Nowak:Pluciennik:1999} (resp., \cite[Cor. 2]{Cui:Hudzik:Pluciennik:2003});
\item[(ii)] $\type(\N)$ $=$ \Isfi\ (resp., II): \textup{\cite[Cor. 2.15]{Cui:Hudzik:Nowak:Pluciennik:1999}}+\textup{\cite[Thm. 2.2.(ii)]{Chen:Hudzik:Wisla:1996}} (resp., \cite[Thm. 2.3]{Hudzik:Zbaszyniak:1997});
\item[(iii)]  $\type(\N)$ $=$ \Isfi\ (resp., II$_1$): \cite[Thm. 2.12]{Cui:Hudzik:Nowak:Pluciennik:1999} (resp., \cite[Thm. 3.9]{Cui:Zhao:2022});
\item[(iv)] follows from equivalence of $\n{\cdot}_\orlicz$ and $\n{\cdot}_{\orlicz}^{\mathrm{O}}$ for any type of $(\X,\mu)$ \cite[Thm. II.2.3]{Luxemburg:1955}, and results quoted in the proof of Proposition \ref{prop.noncommutative.MTNL.norm.geometry.characterisation}.\ref{prop.noncommutative.MTNL.norm.geometry.characterisation.iv}.
\end{enumerate}
\end{proof}

\begin{proposition}
\label{prop.noncommutative.Orlicz.D.ell.Psi}
Let $\orlicz$ be an Orlicz function, let $\N$ be a W$^*$-algebra, either of type \Isfi{}, or of type II$_1$, or of type II$_\infty$, and let $\orlicz,\orlicz^\young\in\triangle_2^0$ (resp., $\triangle_2^\infty$; $\triangle_2$) if $\N$ is of type \Isfi{} (resp., II$_1$; II$_\infty$). Let $\tau$ be a faithful normal semifinite trace on $\N$. Let $\beta\in\,]0,1[$, $\alpha\in\,]0,\infty[$. Let $\Psi\in\pclg(L_\orlicz(\N,\tau),\n{\cdot}_\orlicz)$ be strictly convex on $\efd(\Psi)=L_\orlicz(\N,\tau)$, and consider closed subsets in $L_\orlicz(\N,\tau)$ defined in terms of the topology of $\n{\cdot}_\orlicz$. Then:
\begin{enumerate}[nosep,label=(\roman*)]
\item\label{prop.noncommutative.Orlicz.D.ell.Psi.i} $D_{\ell_\orlicz,\Psi}$ is an information on $\N_\star$;
\item\label{prop.noncommutative.Orlicz.D.ell.Psi.ii} if $\Psi$ is Euler--Legendre, and $\varnothing\neq C\subseteq\N_\star$ is $\ell_\orlicz$-convex and $\ell_\orlicz$-closed, then $C$ is left $D_{\ell_\orlicz,\Psi}$-Chebysh\"{e}v, and $\omega=\LPPP^{D_{\ell_\orlicz,\Psi}}_C(\psi)$ if{}f $\omega$ is the unique solution of
\begin{equation}
\hspace{-1cm}D_{\ell_\orlicz,\Psi}(\phi,\zeta)+D_{\ell_\orlicz,\Psi}(\zeta,\psi)\leq D_{\ell_\orlicz,\Psi}(\phi,\psi)\;\forall(\phi,\psi)\in C\times\N_\star,
\label{eqn.gen.pyth.ineq.Psi.ell.orlicz}
\end{equation}
with $\leq$ replaced by $=$, in `then' case of this `if{}f', if $C$ is $\ell_\orlicz$-affine;
\item\label{prop.noncommutative.Orlicz.D.ell.Psi.iii} if $\orlicz^\young$ is an Orlicz function, $\Psi^\lfdual\in\pclg(L_{\orlicz^\young}(\N,\tau),\n{\cdot}_{\orlicz^\young}^{\mathrm{O}})$ is totally convex, $\varnothing\neq C\subseteq\N_\star$, and $C$ is $(\DG\Psi\circ\ell_\orlicz)$-convex and $(\DG\Psi\circ\ell_\orlicz)$-closed, then $C$ is right $D_{\ell_\orlicz,\Psi}$-Chebysh\"{e}v, and $\omega=\RPPP^{D_{\ell_\orlicz},\Psi}_C(\phi)$ if{}f $\omega$ is the unique solution of
\begin{equation}
\hspace{-1cm}D_{\ell_\orlicz,\Psi}(\phi,\zeta)+D_{\ell_\orlicz,\Psi}(\zeta,\psi)\leq D_{\ell_\orlicz,\Psi}(\phi,\psi)\;\forall(\phi,\psi)\in \N_\star\times C,
\label{eqn.gen.pyth.ineq.Psi.ell.orlicz.right}
\end{equation}
with $\leq$ replaced by $=$, in `then' case of this `if{}f', if $C$ is $(\DG\Psi\circ\ell_\orlicz)$-affine;
\item\label{prop.noncommutative.Orlicz.D.ell.Psi.iv}if $\Psi=\Psi_{\alpha,\beta}=\frac{\beta}{\alpha}\n{\cdot}_\orlicz^{1/\beta}$, then:
\begin{enumerate}[nosep,label=\alph*)]
\item\label{prop.noncommutative.Orlicz.D.ell.Psi.iv.a} the conditions of \ref{prop.noncommutative.Orlicz.D.ell.Psi.i} for $\Psi$ and $\orlicz$ are satisfied;
\item\label{prop.noncommutative.Orlicz.D.ell.Psi.iv.b}if
\begin{equation}
\hspace{-0.5cm}\left\{\begin{array}{ll}
\orlicz\in\SC([0,\orlicz^{\inver}(\frac{1}{2})])\cap\DIFF([0,\orlicz^{\inver}(1)])&\st\type(\N)=\mbox{\Isfi}\\
\orlicz\in\SC(\RR)\cap\DIFF(\RR)&\st\type(\N)=\mbox{II}_1\\
\orlicz\in\SC(\RR)\cap\DIFF(\RR)&\st\type(\N)=\mbox{II}_\infty,
\end{array}\right.
\label{eqn.young.function.conditions.left.right.chebyshev}
\end{equation}
then the conditions of \ref{prop.noncommutative.Orlicz.D.ell.Psi.ii} for $\Psi$ and $\orlicz$ are satisfied;
\item\label{prop.noncommutative.Orlicz.D.ell.Psi.iv.c} if \eqref{eqn.young.function.conditions.left.right.chebyshev} holds, and, additionally, $\orlicz^\young$ is an Orlicz function such that
\begin{equation}
\hspace{-1.4cm}\left\{\begin{array}{ll}
\orlicz^\young\in\SC([0,\varpi_{\orlicz^\young}(1)]),\;\exists u>0\;\;\orlicz((\orlicz^\young)'_+)\geq\frac{1}{2}\\
\;\;\;\;\;\;\;\;\;\st\type(\N)=\mbox{\Isfi}\\
\lim_{u\ra\infty}\frac{\orlicz^\young(u)}{u}=\infty,\;\lim_{u\ra\infty}((\lim_{t\ra\infty}\frac{\orlicz^\young(t)}{t})\ab{u}-\orlicz^\young(u))=\infty\\
\;\;\;\;\;\;\;\;\;\st\type(\N)=\mbox{II}_1\\
\lim_{u\ra\infty}((\lim_{t\ra\infty}\frac{\orlicz^\young(t)}{t})\ab{u}-\orlicz^\young(u))=\infty\\
\;\;\;\;\;\;\;\;\;\st\type(\N)=\mbox{II}_\infty,
\end{array}\right.
\label{eqn.conditions.dual.orlicz.right.chebyshev}
\end{equation}
then the conditions of \ref{prop.noncommutative.Orlicz.D.ell.Psi.iii} for $\Psi$ and $\orlicz$ are satisfied, and $(\DG\Psi\circ\ell_\orlicz)$-closed sets coincide with $\ell_\orlicz$-closed sets;
\end{enumerate}
\item\label{prop.noncommutative.Orlicz.D.ell.Psi.v} if $\Psi$, $\orlicz$, and $\orlicz^\young$ are as in \ref{prop.noncommutative.Orlicz.D.ell.Psi.iv}.\ref{prop.noncommutative.Orlicz.D.ell.Psi.iv.c}, with an additional condition that $\orlicz\in\NFUN$ if $\type(\N)=\mbox{II}_1$, and $\varnothing\neq K\subseteq L_\orlicz(\N,\tau)$ is convex and closed, then $\LPPP^{D_{\Psi}}_K$ is norm-to-norm continuous on $(L_\orlicz(\N,\tau),\n{\cdot}_\orlicz)$.
\end{enumerate}
\end{proposition}
\begin{proof}
\ref{prop.noncommutative.Orlicz.D.ell.Psi.i}--\ref{prop.noncommutative.Orlicz.D.ell.Psi.iii} follow from Propositions \ref{prop.D.ell.psi.properties} and \ref{prop.noncommutative.MTNL.norm.geometry.characterisation}. \ref{prop.noncommutative.Orlicz.D.ell.Psi.iv}--\ref{prop.noncommutative.Orlicz.D.ell.Psi.v} use \cite[Thm. 3.5]{Anderson:1960}: if $(X^{\star},\n{\cdot}_{X^\star})$ is reflexive, strictly convex, and has Radon--Riesz--Shmul'yan property, then $(X,\n{\cdot}_X)$ is Fr\'{e}chet differentiable. \ref{prop.noncommutative.Orlicz.D.ell.Psi.iv}.\ref{prop.noncommutative.Orlicz.D.ell.Psi.iv.a}--\ref{prop.noncommutative.Orlicz.D.ell.Psi.iv}.\ref{prop.noncommutative.Orlicz.D.ell.Psi.iv.b} follow from Propositions \ref{prop.left.pythagorean.info.Psi.alpha.beta} and \ref{prop.Resmerita.cor}, combined with Proposition \ref{prop.noncommutative.MTNL.norm.geometry.characterisation}. In order to identify the conditions \eqref{eqn.conditions.dual.orlicz.right.chebyshev} as sufficient in \ref{prop.noncommutative.Orlicz.D.ell.Psi.iv}.\ref{prop.noncommutative.Orlicz.D.ell.Psi.iv.c} case, we use Propositions \ref{prop.nc.Orlicz.spaces.Banach.duality} and \ref{prop.noncommutative.Orlicz.norm.geometry.characterisation}. Since Proposition \ref{prop.noncommutative.MTNL.norm.geometry.characterisation} is missing the characterisation of Radon--Riesz--Shmul'yan property for $\type(\N)=\mbox{II}_\infty$, we additionally use \cite[Thms. 3, 11]{Hudzik:Kowalewski:Lewicki:2006}, stating that, for any Orlicz $\orlicz$ and for any nonatomic $(\X,\mu)$, $(L_\orlicz(\X,\mu),\n{\cdot}_\orlicz^{\mathrm{O}})$ (has Radon--Riesz--Shmul'yan property and is reflexive) if{}f (it is strictly convex and reflexive). For $\type(\N)=\mbox{II}$, the conditions used in \ref{prop.noncommutative.Orlicz.D.ell.Psi.iv}.\ref{prop.noncommutative.Orlicz.D.ell.Psi.iv.c} do not include $\orlicz^\young\in\SC(\RR)$, since it is equivalent with $\orlicz\in\DIFF(\RR)$, and the latter is already assumed in \ref{prop.noncommutative.Orlicz.D.ell.Psi.iv}.\ref{prop.noncommutative.Orlicz.D.ell.Psi.iv.b}. 
\end{proof}

\begin{proposition}\label{cor.commutative.Orlicz}
Let $\orlicz$ be an Orlicz function such that $\orlicz(1)=1$, and there exist $t,s\in\RR^+\setminus\{0\}$ such that $t<s$, $u\mapsto\frac{\orlicz^{\inver}(u)}{u^t}$ is nondecreasing, and $u\mapsto\frac{\orlicz^{\inver}(u)}{u^s}$ is nonincreasing. Let $(\X,\mu)$ be a measure space, such that one of the following conditions holds:
\begin{enumerate}[nosep,label=\alph*)]
\item\label{cor.commutative.Orlicz.a} $(\X,\mu)$ is purely atomic, $\mu(\X)=\infty$, $\orlicz\in\triangle_2^0\cap\SC([0,\orlicz^{\inver}(\frac{1}{2})])\cap\DIFF([0,$ $\orlicz^{\inver}(1)])$, $\orlicz^\young\in\triangle_2^0$;
\item\label{cor.commutative.Orlicz.b}$(\X,\mu)$ is atomless, $\mu(\X)<\infty$, $\orlicz\in\triangle_2^\infty\cap\SC(\RR)\cap\DIFF(\RR)$, $\orlicz^\young\in\triangle_2^\infty$;
\item\label{cor.commutative.Orlicz.c} $(\X,\mu)$ is atomless, $\mu(\X)=\infty$, $\orlicz\in\triangle_2\cap\SC(\RR)\cap\DIFF(\RR)$, $\orlicz^\young\in\triangle_2$.
\end{enumerate}
Let $\Psi=\Psi_{\beta,\beta}=\n{\cdot}_\orlicz^{1/\beta}:L_\orlicz(\X,\mu)\ra\RR^+$, $\beta\in\,]0,1[$. Let $\varnothing\neq C\subseteq B(L_1(\X,\mu),\n{\cdot}_1)$ (resp., $\varnothing\neq C\subseteq S(L_1(\X,\mu),\n{\cdot}_1)$) be $\ell_\orlicz$-convex and closed. Then:
\begin{enumerate}[nosep,label=(\roman*)]
\item\label{cor.commutative.Orlicz.i} $\LPPP^{D_{\ell_\orlicz,\Psi_{\beta,\beta}}}_C$ satisfies \eqref{eqn.gen.pyth.ineq.Psi.ell.orlicz}, and is norm-to-norm continuous on $B(L_1(\X,\mu),$ $\n{\cdot}_1)$ (resp., $S(L_1(\X,$ $\mu),\n{\cdot}_1)$);
\item\label{cor.commutative.Orlicz.ii} $D_{\ell_\orlicz,\Psi_{\varphi_{\beta,\beta}}}:(L_1(\X,\mu))^+\times (L_1(\X,\mu))^+\ra\RR^+$ is given by $\forall\omega,\phi\in(L_1(\X,\mu))^+$
\begin{align}
D_{\ell_\orlicz,\Psi_{\varphi_{\beta,\beta}}}(\omega,\phi)=&\n{\orlicz^{\inver}(\omega)}_\orlicz^{1/\beta}+\frac{1-\beta}{\beta}\n{\orlicz^{\inver}(\phi)}_\orlicz^{1/\beta}\nonumber\\&-\frac{1}{\beta}\n{\orlicz^{\inver}(\phi)}_\orlicz^{1/\beta-1}\frac{\int\mu\orlicz^{\inver}(\omega)\orlicz'\left(\frac{\orlicz^{\inver}(\phi)}{\n{\orlicz^{\inver}(\phi)}_\orlicz}\right)}{\int\mu\orlicz^{\inver}(\phi)\orlicz'\left(\frac{\orlicz^{\inver}(\phi)}{\n{\orlicz^{\inver}(\phi)}_\orlicz}\right)},
\label{eqn.D.orlicz.beta.beta}
\end{align}
where $\orlicz'(t):=\frac{\dd \orlicz(t)}{\dd t}>0$ $\forall t>0$;
\item\label{cor.commutative.Orlicz.iii} in particular, for $\bar{\orlicz}(\omega,\phi):=\int\mu\orlicz^{\inver}(\omega)\orlicz'(\orlicz^{\inver}(\phi))$,
\begin{equation}
\hspace{-0.3cm}D_{\ell_\orlicz,\Psi_{\varphi_{\beta,\beta}}}(\omega,\phi)=\frac{1}{\beta}\left(1-\frac{\bar{\orlicz}(\omega,\phi)}{\bar{\orlicz}(\phi,\phi)}\right)\;\;\forall\omega,\phi\in (S(L_1(\X,\mu)),\n{\cdot}_1)^+.
\end{equation}
\end{enumerate}
\end{proposition}
\begin{proof}
\begin{enumerate}[nosep]
\item[(i)] For any Orlicz $\orlicz$, if $\orlicz(1)=1$, and there exist $t,s\in\RR^+\setminus\{0\}$ such that $t<s$, $u\mapsto\frac{\orlicz^{\inver}(u)}{u^t}$ is nondecreasing, and $u\mapsto\frac{\orlicz^{\inver}(u)}{u^s}$ is nonincreasing, then the Kaczmarz map between unit balls and unit spheres of $(L_1(\X,\mu),\n{\cdot}_1)$ and $(L_\orlicz(\X,\mu),\n{\cdot}_\orlicz)$ is uniformly homeomorphic \cite[Thm. 2.4, p. 199]{Delpech:2005} (=\cite[Thm. 4.5, p. 32]{Delpech:2005:PhD}).\footnote{More precisely, this theorem asserts Lipschitz--H\"{o}lder continuity for Kaczmarz maps $\ell_{\orlicz_1,\orlicz_2}$ with $\orlicz_1,\orlicz_2\in\NFUN$. Yet, given any Orlicz functions $\orlicz_1$ and $\orlicz_2$, the assumption $\orlicz_1,\orlicz_2\in\NFUN$ is not used anywhere in the proof of this theorem (the only place implicitly applying it is a proof of a formula $\n{\frac{\orlicz_1((\ab{x}(\cdot)+\ab{y}(\cdot))/4)}{(\ab{x}(\cdot)+\ab{y}(\cdot))/4}}_{\orlicz_1^\young}\leq1$, but the latter holds trivially for $\orlicz_1=\ab{\cdot}$).} The rest follows by a conjunction of Propositions \ref{prop.left.pythagorean.info.Psi.alpha.beta}.\ref{prop.left.pythagorean.info.Psi.alpha.beta.iv}, \ref{prop.noncommutative.MTNL.norm.geometry.characterisation}, and \ref{prop.noncommutative.Orlicz.D.ell.Psi}.
\item[(ii)] By \cite[Lem. 3]{Grzaslewicz:Hudzik:1992} (cf. also \cite[Eqn. (18.29)]{Krasnoselskii:Rutickii:1958}, \cite[\S3]{Lumer:1963}, \cite[Eqn. (10)]{Rao:1965:I}), if $\n{\cdot}_\orlicz$ is Gateaux differentiable, then
\begin{equation}
\DG\n{x}_\orlicz=\frac{\orlicz'\left(\frac{x}{\n{x}_\orlicz}\right)}{\int\mu\frac{x}{\n{x}_\orlicz}\orlicz'\left(\frac{x}{\n{x}_\orlicz}\right)}\;\;\forall x\in L_\orlicz(\X,\mu)\setminus\{0\}.
\label{eqn.DG.of.MTN.norm}
\end{equation}
\item[(iii)] Follows from \eqref{eqn.D.orlicz.beta.beta} by a direct calculation.
\end{enumerate}
\end{proof}

\subsection{Generalised spin factors}

\begin{proposition}\label{prop.gen.spin.factor.D.alpha.beta.gamma}
Let $(V,\n{\cdot}_V)$ be a radially compact base normed space defined by $V=X\oplus\RR$, where $(X,\n{\cdot}_X)$ is a reflexive Banach space, and
\begin{equation}
\forall\phi=(x,\lambda)\in V\;\;\left\{\begin{array}{l}
\phi\geq0\;:\iff\;\lambda\geq\n{x}_X\\
\n{\phi}_V:=\max\{\ab{\lambda},\n{x}_X\}.
\end{array}\right.
\label{eqn.generalised.spin.factor.order.norm}
\end{equation}
Let $\beta\in\,]0,1[$, $\alpha\in\,]0,\infty[$, and let
\begin{equation}
\ell_{/\RR}:(S(V,\n{\cdot}_V))^+\ni\phi=:(x,1)\mapsto x\in B(X,\n{\cdot}_X).
\end{equation}
Then:
\begin{enumerate}[nosep,label=(\roman*)]
\item\label{prop.gen.spin.factor.D.alpha.beta.gamma.i} $(V,\n{\cdot}_V)$ satisfies spectral duality condition of \textup{\cite[Def. (p. 55)]{Alfsen:Shultz:1976}} if{}f $(X,\n{\cdot}_X)$ is strictly convex and Gateaux differentiable if{}f $\Psi_{\alpha,\beta}:X\ra\RR^+$ is Euler--Legendre with respect to $\n{\cdot}_X$;
\item\label{prop.gen.spin.factor.D.alpha.beta.gamma.ii}$\ell_{/\RR}$ is a norm-to-norm homeomorphism;
\item\label{prop.gen.spin.factor.D.alpha.beta.gamma.iii} if any of the equivalent conditions in \ref{prop.gen.spin.factor.D.alpha.beta.gamma.i} is satisfied, then:
\begin{enumerate}[nosep,label=\alph*)]
\item\label{prop.gen.spin.factor.D.alpha.beta.gamma.iii.a} $D_{\ell_{/\RR},\Psi_{\alpha,\beta}}:(S(V,\n{\cdot}_V))^+\times(S(V,\n{\cdot}_V))^+\ra\RR^+$ is an information on $(S(V,\n{\cdot}_V))^+$;
\item\label{prop.gen.spin.factor.D.alpha.beta.gamma.iii.b} if $\varnothing\neq C\subseteq(S(V,\n{\cdot}_V))^+$ is $\ell_{/\RR}$-convex and closed, then $C$ is left $D_{\ell_{/\RR},\Psi_{\alpha,\beta}}$-Chebysh\"{e}v, and $\omega=\LPPP^{D_{\ell_{/\RR},\Psi_{\alpha,\beta}}}_C(\psi)$ if{}f $\omega$ is the unique solution of $\forall(\phi,\psi)\in C\times(S(V,\n{\cdot}_V))^+$
\begin{equation}
D_{\ell_{/\RR},\Psi_{\alpha,\beta}}(\phi,\zeta)+D_{\ell_{/\RR},\Psi_{\alpha,\beta}}(\zeta,\psi)\leq D_{\ell_{/\RR},\Psi_{\alpha,\beta}}(\phi,\psi),
\end{equation}
with $\leq$ replaced by $=$, in `then' case of this `if{}f', if $C$ is $\ell_{/\RR}$-affine;
\item\label{prop.gen.spin.factor.D.alpha.beta.gamma.iii.c} if $(X,\n{\cdot}_X)$ is Fr\'{e}chet differentiable, then:
\begin{enumerate}[nosep]
\item[1)] if $C$ is as in \ref{prop.gen.spin.factor.D.alpha.beta.gamma.iii.b} and $(X,\n{\cdot}_X)$ has Radon--Riesz--Shmul'yan property, then $\LPPP^{D_{\ell_{/\RR},\Psi_{\alpha,\beta}}}_C$ is norm-to-norm continuous on $(S(V,\n{\cdot}_V))^+$;
\item[2)] if $\varnothing\neq\widetilde{C}\subseteq(S(V,\n{\cdot}_V))^+$ and $\widetilde{C}$ is $(\DG\Psi_{\alpha,\beta}\circ\ell_{/\RR})$-convex and closed, then it is right $D_{\ell_{/\RR},\Psi_{\alpha,\beta}}$-Chebysh\"{e}v, and $\omega=\RPPP^{D_{\ell_{/\RR},\Psi_{\alpha,\beta}}}_{\widetilde{C}}(\phi)$ if{}f $\omega$ is the unique solution of $\forall(\phi,\psi)\in(S(V,\n{\cdot}_V))^+\times\widetilde{C}$
\begin{equation}
D_{\ell_{/\RR},\Psi_{\alpha,\beta}}(\phi,\zeta)+D_{\ell_{/\RR},\Psi_{\alpha,\beta}}(\zeta,\psi)\leq D_{\ell_{/\RR},\Psi_{\alpha,\beta}}(\phi,\psi);
\end{equation}
with $\leq$ replaced by $=$, in `then' case of this `if{}f', if $\widetilde{C}$ is $(\DG\Psi_{\alpha,\beta}\circ\ell_{/\RR})$-affine.
\end{enumerate}
\end{enumerate}
\end{enumerate}
\end{proposition}
\begin{proof}
According to \cite[Thm. 1]{Berdikulov:Odilov:1995} (recently independently rediscovered in \cite[Thm. 6.6]{Jencova:Pulmannova:2021}), $(V=X\oplus\RR,\n{\cdot}_V)$ defined by \eqref{eqn.generalised.spin.factor.order.norm} satisfies spectral duality condition if{}f $(X,\n{\cdot}_X)$ is reflexive, strictly convex, and Gateaux differentiable. Combining this with Proposition \ref{prop.BBC.lemma} and Remark \ref{remark.BCC.Resmerita.extension} gives \ref{prop.gen.spin.factor.D.alpha.beta.gamma.i}. \ref{prop.gen.spin.factor.D.alpha.beta.gamma.ii} follows from the fact that $\n{\phi_1-\phi_2}_V=\n{x_1-x_2}_X$ $\forall\phi_1=(x_1,1),\phi_2=(x_2,1)\in(S(V,\n{\cdot}_V))^+$. The rest follows from Proposition \ref{prop.left.pythagorean.info.Psi.alpha.beta}.
\end{proof}

\begin{definition} \textup{\cite[Def. 4]{Berdikulov:Odilov:1995}} If $(V,\n{\cdot}_V)$ satisfies the conditions of Proposition \ref{prop.gen.spin.factor.D.alpha.beta.gamma}.\ref{prop.gen.spin.factor.D.alpha.beta.gamma.i}, then $(V,\n{\cdot}_V)^\star$ is called a \df{generalised spin factor}.
\end{definition}

\begin{remark}\label{remark.spin.factors.are.JBW.algebras}\textup{\cite[Thm. 7.1]{Stoermer:1966:Jordan}} If $(X,\n{\cdot}_X)$ is a Hilbert space and $\dim X\geq 2$, then $(V,\n{\cdot}_V)^\star$ coincide with type I$_2$ JBW-algebras.
\end{remark}

\subsection{Lozanovski\u{\i} factorisation maps}

\begin{proposition}\label{prop.lozanovskii.D.ell.Psi}
Let $(E(\X,\mu),\n{\cdot}_{E(\X,\mu)})$ be a uniformly convex and uniformly Fr\'{e}chet differentiable Banach function space over a localisable measure space, let ${\ell_E}^\inver:S(E(\X,\mu),\n{\cdot}_{E(\X,\mu)})\ni x\mapsto\ab{j(x)}x\in S(L_1(\X,\mu),\n{\cdot}_1)$, let $\Psi\in\pclg(E(\X,\mu),\n{\cdot}_{E(\X,\mu)})$ be strictly convex on $\efd(\Psi)=E(\X,\mu)$, let $\beta\in\,]0,1[$, $\alpha\in\,]0,\infty[$, and $\varnothing\neq C\subseteq S(L_1(\X,\mu),\n{\cdot}_1)$. Then:
\begin{enumerate}[nosep,label=(\roman*)]
\item\label{prop.lozanovskii.D.ell.Psi.i} ${\ell_E}^\inver$ is a uniformly continuous homeomorphism, with
\begin{equation}
\ell_E(\phi)=\RPPP^{\bar{D}_1}_{S(E(\X,\mu),\n{\cdot}_{E(\X,\mu)})}(\phi)\;\;\forall\phi\in S(L_1(\X,\mu),\n{\cdot}_1)\cap L_\infty(\X,\mu),
\label{eqn.lozanovskii.ell.RPPP.D.1}
\end{equation}
where $\forall(x,y)\in S(L_1(\X,\mu),\n{\cdot}_1)\cap L_\infty(\X,\mu)\times S(E(\X,\mu),\n{\cdot}_{E(\X,\mu)})$
\begin{equation}
\bar{D}_1(x,y):=\sgn(x)(\ab{x}\log(\textstyle\frac{\ab{x}}{\ab{y}}));
\end{equation}
\item\label{prop.lozanovskii.D.ell.Psi.ii} $D_{\ell_E,\Psi}$ is an information on $S(L_1(\X,\mu),\n{\cdot}_1)$;
\item\label{prop.lozanovskii.D.ell.Psi.iii} if $\Psi$ is Euler--Legendre and $C$ is $\ell_E$-convex and closed, then $C$ is $D_{\ell_E,\Psi}$-Chebysh\"{e}v, and $\omega=\LPPP^{D_{\ell_E,\Psi}}_C(\psi)$ if{}f $\omega$ is a unique solution of $\forall(\phi,\psi)\in C\times S(L_1(\X,\mu),\n{\cdot}_1)$
\begin{equation}
D_{\ell_E,\Psi}(\phi,\zeta)+D_{\ell_E,\Psi}(\zeta,\psi)\leq D_{\ell_E,\Psi}(\phi,\psi),
\end{equation}
with $\leq$ replaced by $=$, in `then' case of `if{}f', if $C$ is $\ell_E$-affine;
\item\label{prop.lozanovskii.D.ell.Psi.iv} if $\Psi^\lfdual\in\pclg((E(\X,\mu))^\star,{\n{\cdot}_{E(\X,\mu)}}^\star)$ is totally convex and $C$ is $(\DG\Psi\circ\ell_E)$-convex and $(\DG\Psi\circ\ell_E)$-closed, then $C$ is right $D_{\ell_E,\Psi}$-Chebysh\"{e}v, and $\omega=\RPPP^{D_{\ell_E,\Psi}}_C(\phi)$ if{}f $\omega$ is a unique solution of $\forall(\phi,\psi)\in S(L_1(\X,\mu),\n{\cdot}_1)\times C$
\begin{equation}
D_{\ell_E,\Psi}(\phi,\zeta)+D_{\ell_E,\Psi}(\zeta,\psi)\leq D_{\ell_E,\Psi}(\phi,\psi),
\end{equation}
with $\leq$ replaced by $=$, in `then' case of `if{}f', if $C$ is $(\DG\Psi\circ\ell_E)$-affine;
\item\label{prop.lozanovskii.D.ell.Psi.v} if $\Psi=\Psi_{\alpha,\beta}=\frac{\beta}{\alpha}{\n{\cdot}_{E(\X,\mu)}}^{1/\beta}$, then:
\begin{enumerate}[nosep,label=\alph*)]
\item\label{prop.lozanovskii.D.ell.Psi.v.a} the assumptions, and thus the conclusions, of \ref{prop.lozanovskii.D.ell.Psi.ii}--\ref{prop.lozanovskii.D.ell.Psi.iv} hold for $D_{\ell_E,\Psi}=D_{\ell_E,\Psi_{\alpha,\beta}}$, and the $(\DG\Psi_{\alpha,\beta}\circ\ell_E)$-closed sets are closed;
\item\label{prop.lozanovskii.D.ell.Psi.v.b}$\LPPP^{D_{\ell_E,\Psi_{\alpha,\beta}}}_C$ is norm-to-norm continuous on $S(L_1(\X,\mu),\n{\cdot}_1)$;
\item\label{prop.lozanovskii.D.ell.Psi.v.c}$\forall\phi,\psi\in S(L_1(\X,\mu),\n{\cdot}_1)\cap L_\infty(\X,\mu)$
\begin{equation}
\hspace{-1.2cm}D_{\ell_E,\Psi_{\alpha,\beta}}(\phi,\psi)=\textstyle\frac{1}{\alpha}(1-\duality{\RPPP^{\bar{D}_1}_{S(E(\X,\mu),\n{\cdot}_{E(\X,\mu)})}(\phi),j\circ\RPPP^{\bar{D}_1}_{S(E(\X,\mu),\n{\cdot}_{E(\X,\mu)})}(\psi)}_{E(\X,\mu)\times(E(\X,\mu))^\star}).
\end{equation}
\end{enumerate}
\end{enumerate}
\end{proposition}
\begin{proof}
\begin{enumerate}[nosep]
\item[(i)] This was proved in \cite[Prop. 2.6]{Odell:Schlumprecht:1994} (resp., \cite[Props. 2.8, 2.9]{Chaatit:1995}) for atomic infinite (resp., countably finite) $(\X,\mu)$ and $\widetilde{D}_1(x,y):=-\sgn(x)\ab{x}\log\ab{y}$. The restriction to countably finite $(\X,\mu)$ was observed to be unnecessary in \cite[Thm. 12]{Raynaud:1997}. The extension from $\widetilde{D}_1$ to $\bar{D}_1$ was observed in \cite[\S5]{ChavezDominguez:2023}.
\item[(ii)--(v)] Follows from \ref{prop.lozanovskii.D.ell.Psi.i} and Proposition \ref{prop.left.pythagorean.info.Psi.alpha.beta}.
\end{enumerate}
\end{proof}

\begin{proposition}\label{prop.nc.lozanovskii.D.ell.Psi}
\sloppy Let $\N=\BH$ be a W$^*$-algebra of type I$_n$ with $n\in\NN$, let $(E(\N,\tr_\H),\n{\cdot}_{E(\N,\tr_\H)})$ be a rearrangement invariant space (i.e. a unitary invariant ideal of compact operators on an $n$-dimensional Hilbert space $\H$ \textup{\cite[\S\S6--7, Rem. 4]{vonNeumann:1937:Tomsk}}) that is uniformly convex and uniformly Fr\'{e}\-chet differentiable, let ${\ell_E}^\inver:S(E(\N,\tr_\H),\n{\cdot}_{E(\N,\tr_\H)})\ni x\mapsto\ab{j(x)}x\in S(\N_\star,\n{\cdot}_1)$, let $\Psi\in\pclg(E(\N,\tr_\H),\n{\cdot}_{E(\N,\tr_\H)})$ be strictly convex on $\efd(\Psi)=E(\N,\tr_\H)$, let $\beta\in\,]0,1[$, $\alpha\in\,]0,\infty[$, and $\varnothing\neq C\subseteq(S(\N_\star,\n{\cdot}_1))^+$. Then:
\begin{enumerate}[nosep,label=(\roman*)]
\item\label{prop.nc.lozanovskii.D.ell.Psi.i}${\ell_E}^\inver$ is a uniformly continuous homeomorphism on $(S(E(\N,\tr_\H),\n{\cdot}_{E(\N,\tr_\H)}))^+$, with
\begin{equation}
\ell_E(\phi)=\RPPP^{\bar{D}_1}_{(S(E(\N,\tr_\H),\n{\cdot}_{E(\N,\tr_\H)}))^+}(\phi)\;\;\forall\phi\in(S(\N_\star,\n{\cdot}_1))^+,
\label{eqn.nc.lozanovskii.ell.RPPP.D.1}
\end{equation}
where $\bar{D}_1(x,y):=\ab{x}\log\ab{x}-\ab{x}\log\ab{y}$ $\forall(x,y)\in(S(\N_\star,\n{\cdot}_1))^+\times(S(E(\N,\tr_\H),\n{\cdot}_{E(\N,\tr_\H)}))^+$.
\item\label{prop.nc.lozanovskii.D.ell.Psi.ii}all statements of Proposition \ref{prop.lozanovskii.D.ell.Psi}.\ref{prop.lozanovskii.D.ell.Psi.ii}--\ref{prop.lozanovskii.D.ell.Psi.v}.\ref{prop.lozanovskii.D.ell.Psi.v.b} hold, under replacement of $((E(\X,\mu))^\star,{\n{\cdot}_{E(\X,\mu)}}^\star)$ by $((E(\N,\tr_\H))^\star,{\n{\cdot}_{E(\N,\tr_\H)}}^\star)$, $S(L_1(\X,\mu),\n{\cdot}_1)$ by $(S(\N_\star,\n{\cdot}_1))^+$, and $\Psi_{\alpha,\beta}=\frac{\beta}{\alpha}{\n{\cdot}_{E(\X,\mu)}}^{1/\beta}$ by $\Psi_{\alpha,\beta}=\frac{\beta}{\alpha}{\n{\cdot}_{E(\N,\tr_\H)}}^{1/\beta}$;
\item\label{prop.nc.lozanovskii.D.ell.Psi.iii}if $\Psi=\Psi_{\alpha,\beta}=\frac{\beta}{\alpha}{\n{\cdot}_{E(\N,\tr_\H)}}^{1/\beta}$ and $\phi,\psi\in(S(\N_\star,\n{\cdot}_1))^+$, then
\begin{equation}
\hspace{-0.4cm}D_{\ell_E,\Psi_{\alpha,\beta}}(\phi,\psi)=\textstyle\frac{1}{\alpha}(1-\duality{\RPPP^{\bar{D}_1}_{K}(\phi),j\circ\RPPP^{\bar{D}_1}_{K}(\psi)}_{E(\N,\tr_\H)\times(E(\N,\tr_\H))^\star}),
\end{equation}
where $K:=(S(E(\N,\tr_\H),\n{\cdot}_{E(\N,\tr_\H)}))^+$.
\end{enumerate}
\end{proposition}
\begin{proof}
\begin{enumerate}[nosep]
\item[(i)] This was proved in \cite[Props. 5.6, 5.7, Lem. 5.8]{ChavezDominguez:2023}.
\item[(ii)--(iii)] Follows from \ref{prop.nc.lozanovskii.D.ell.Psi.i} and Proposition \ref{prop.left.pythagorean.info.Psi.alpha.beta}.
\end{enumerate}
\end{proof}

\begin{proposition}\label{prop.when.lozanovskii.equals.kaczmarz}
\sloppy Let $(\X,\mu)$ be nonatomic, let $\orlicz$ be a finite Young function, let $(X,\n{\cdot}_X)=(L_\orlicz(\X,\mu),\n{\cdot}_\orlicz)$ be Gateaux differentiable, and such that the Lozanovski\u{\i} factorisation map $\ell_X:S(L_1(\X,\mu,\n{\cdot}_1)\ra S(L_\orlicz(\X,\mu),\n{\cdot}_\orlicz)$ is a bijection. Then $\ell_X=\ell_\orlicz$ if{}f
\begin{equation}
\ab{\frac{\orlicz'(x)}{\int_\X\mu(\xx)\orlicz(x(\xx))}}=\frac{\orlicz(\ab{x})}{\ab{x}}\;\;\forall x\in S(L_\orlicz(\X,\mu),\n{\cdot}_\orlicz).
\label{eqn.when.lozanovskii.equals.kaczmarz}
\end{equation}
\end{proposition}
\begin{proof}
From \cite[p. 200]{Asplund:1967} for $\varphi(t)=t$ we obtain $j(x)=\n{x}_X\DG\n{x}_X$. On the other hand, ${\ell_X}^\inver(x)=\ab{j(x)}x$. Hence, ${\ell_X}^\inver(x)=\n{x}_X\ab{\DG\n{x}_X}x$. Hence, from \eqref{eqn.DG.of.MTN.norm} it follows that
\begin{equation}
{\ell_X}^\inver=\n{x}_\orlicz\ab{\frac{\orlicz'\left(\frac{x}{\n{x}_\orlicz}\right)}{\int\mu\frac{x}{\n{x}_\orlicz}\orlicz'\left(\frac{x}{\n{x}_\orlicz}\right)}}x.
\end{equation}
So, the equation $\orlicz(\ab{x})\frac{x}{\ab{x}}={\ell_\orlicz}^\inver(x)={\ell_X}^\inver(x)$, evaluated at any $x\in S(L_\orlicz(\X,\mu),\n{\cdot}_\orlicz)$, is equivalent with \eqref{eqn.when.lozanovskii.equals.kaczmarz}.
\end{proof}

\section*{Acknowledgments}

\noindent This work was partially founded by MAB/2018/5 grant of Foundation for Polish Science, 2015/18/E/ ST2/00327 and 2021/42/A/ST2/00356 grants of Polish National Center of Science,  IMPULZ IM-2023-79 and VEGA 2/0164/25 grants of Slovak Academy of Sciences, APVV-22-0570 grant of Slovak Research and Development Agency, FellowQUTE 2024-02 program of Slovak National Center for Quantum Technologies, and by Perimeter Institute for Theoretical Physics. Research at Perimeter Institute is supported by the Government of Canada through the Department of Innovation, Science and Economic Development and by the Province of Ontario through the Ministry of Colleges and Universities. Part of this research was conducted at Department of Mathematical Informatics, Graduate School of Information Science, Nagoya University, on an academic leave from University of Gda\'{n}sk during the spring of 2020. I thank: Fran\-ce\-sco Buscemi, Lucien Hardy, Ravi Kunjwal, Jerzy Lewandowski, and Marcin Marciniak for hospitality; Micha{\l} Eckstein, Pawe{\l} Horodecki, and Hamed Mohammady for support.

\section*{References}
\addcontentsline{toc}{section}{References}
{\small
{Unless explicitly stated otherwise, all citations refer to first editions of the corresponding texts in their original language. Symbol * in front of a bibliographic item indicates that I have not seen this work. Cyrillic Russian names and titles were transliterated from original using the following system (which is bijective due to the lack of {\fontencoding{T2A}\selectfont ыа} and {\fontencoding{T2A}\selectfont ыу} combinations): {\fontencoding{T2A}\selectfont ц} = c, {\fontencoding{T2A}\selectfont ч} = ch, {\fontencoding{T2A}\selectfont х} = kh, {\fontencoding{T2A}\selectfont ж} = zh, {\fontencoding{T2A}\selectfont ш} = sh, {\fontencoding{T2A}\selectfont щ} = \v{s}, {\fontencoding{T2A}\selectfont и} = i, {\fontencoding{T2A}\selectfont й} = \u{\i}, i = \={\i}, {\fontencoding{T2A}\selectfont ы} = y, {\fontencoding{T2A}\selectfont ю} = yu, {\fontencoding{T2A}\selectfont я} = ya, {\fontencoding{T2A}\selectfont ё} = \"{e}, {\fontencoding{T2A}\selectfont э} = \`{e}, {\fontencoding{T2A}\selectfont ъ} = `, {\fontencoding{T2A}\selectfont ь} = ', and analogously for capitalised letters, with an exception of {\fontencoding{T2A}\selectfont Х} = H at the beginnings of words. When\-ev\-er possible, Chinese Mandarin (resp., Cantonese) names and titles were nonbijectively romanised from original, using p\={\i}ny\={\i}n (resp., toneless Yale).}

{\scriptsize
\begingroup
\raggedright
\renewcommand\refname{

\endgroup     
}}
\end{document}